\documentclass[11pt]{article}
\usepackage{amsmath}
\usepackage{graphicx}
\usepackage{latexsym}
\usepackage{amssymb}
\usepackage{amsbsy}
\usepackage{color}
\usepackage{lineno}
\usepackage{ulem}
\usepackage{multirow}
\usepackage{rotating}
\DeclareSymbolFont{msbm}{U}{msb}{m}{n}
\DeclareMathSymbol{\C}{\mathalpha}{msbm}{'103}
\DeclareMathSymbol{\R}{\mathalpha}{msbm}{'122}
\DeclareMathSymbol{\Z}{\mathalpha}{msbm}{'132}
\DeclareMathSymbol{\N}{\mathalpha}{msbm}{'116}

\newtheorem{remark}{Remark}

\def\be{\begin{equation}}
\def\ee{\end{equation}}
\def\bea{\begin{eqnarray}}
\def\ba{\begin{array}{l}\displaystyle}
\def\eea{\end{eqnarray}}
\def\ea{\end{array}}
\def\E{{\cal E}}
\def\ca{\\[+0.3cm]\displaystyle}

\begin{document}
\title{ \Large
 Towards an ultra efficient kinetic scheme\\ Part I: basics on the BGK equation \thanks{This work was supported by
the ANR Blanc project BOOST and the ANR JCJC project ALE INC(ubator) 3D}}
\author{Giacomo Dimarco\thanks{Universit\'{e} de Toulouse; UPS, INSA, UT1, UTM;
CNRS, UMR 5219; Institut de Math\'{e}matiques de Toulouse; F-31062
Toulouse, France. ({\tt giacomo.dimarco@math.univ-toulouse.fr}).}
\and Rapha\"{e}l Loubere\thanks{Universit\'{e} de Toulouse; UPS,
INSA, UT1, UTM; CNRS, UMR 5219; Institut de Math\'{e}matiques de
Toulouse; F-31062 Toulouse, France. ({\tt
raphael.loubere@math.univ-toulouse.fr}).} }
\date{\today}
\maketitle
\begin{abstract}
In this paper we present a new ultra efficient numerical method for
solving kinetic equations. In this preliminary work, we present the
scheme in the case of the BGK relaxation operator. The scheme, being
based on a splitting technique between transport and collision, can
be easily extended to other collisional operators as the Boltzmann
collision integral or to other kinetic equations such as the Vlasov
equation. The key idea, on which the method relies, is to solve the
collision part on a grid and then to solve exactly the transport
linear part by following the characteristics backward in time. The
main difference between the method proposed and semi-Lagrangian
methods is that here we do not need to reconstruct the distribution
function at each time step. This allows to tremendously reduce the
computational cost of the method and it permits for the first time,
to the author's knowledge, to compute solutions of full six
dimensional kinetic equations on a single processor laptop machine.
Numerical examples, up to the full three dimensional case, are
presented which validate the method and assess its efficiency
in 1D, 2D and 3D.
\end{abstract}


\section{Introduction}

The kinetic equations provide a mesoscopic description of gases and
more generally of particle systems. In many applications, the
correct physical solution for a system far from thermodynamical
equilibrium, such as rarefied gases or plasmas, requires the
resolution of a kinetic equation \cite{cercignani}. However, the
numerical simulation of these equations with deterministic
techniques presents several drawbacks due to the large dimension of
the problem. The distribution function depends on seven independent
variables: three coordinates in physical space, three coordinates in
velocity space and the time. As a consequence, probabilistic
techniques such as Direct Simulation Monte Carlo (DSMC) methods
\cite{bird, Cf, CPima, Nanbu80} are extensively used in real
situations due to their large flexibility and low computational cost
compared to finite volume, finite difference or spectral methods for
kinetic equations \cite{Filbet2, Filbet, Mieussens, Pal,
Pieraccini}. On the other hand, DSMC solutions are affected by large
fluctuations. Moreover, in non stationary situations it is
impossible to use time averages to reduce these fluctuations and
this leads to, either poorly accurate solutions, or again to
computationally expensive simulations.

For this reason, many different works have been dedicated to reduce
some of the disadvantages of Monte Carlo methods. We quote \cite{Cf}
for an overview on efficient and low variance Monte Carlo methods.
For applications of variance reduction techniques to kinetic
equation let us remind to the works of Homolle and Hadjiconstantinou
\cite{Hadji} and \cite{Hadji1}. We mention also the work of Boyd and
Burt \cite{Boyd} and of Pullin \cite{Pullin78} who developed a low
diffusion particle method for simulating compressible inviscid
flows. We finally quote the works of Dimarco and Pareschi
\cite{dimarco2, dimarco3} and of Degond, Dimarco and Pareschi
\cite{dimarco1} who constructed efficient and low variance methods
for kinetic equations in transitional and general regimes.

In this work, we consider the development of a new deterministic
method to solve kinetic equations. In particular, we focus on the
development of efficient techniques for the discretization of the
linear transport part of these equations. The proposed method is
based on the so-called discrete velocity models (DVM)
\cite{Mieussens} and on the semi Lagrangian approach \cite{CrSon,
CrSon1}. The DVM models are obtained by discretizing the velocity
space into a set of fixed discrete velocities \cite{bobylev,
Mieussens, Pal, Pal1}. As a result of this discretization, the
original kinetic equation is then represented as a set of linear
transport equations plus an interaction term which couples all the
equations. In order to solve the resulting set of equations, the
most common strategy consists in an operator splitting strategy
\cite{Des}: The solution in one time step is obtained by the
sequence of two stages. First one integrates the space homogeneous
equations and then, in the second stage, the transport equation
using the output of the previous step as initial condition. More
sophisticated splitting techniques can be employed, which permits to
obtain high order in time discretizations of the kinetic equations
as for instance the Strang splitting method \cite{Strang}. In any
case, the resulting method is very simple and robust but the main
drawback is again the excessive computational cost. It is a matter
of fact that the numerical solution through such microscopic models
and deterministic schemes remains nowadays too expensive especially
in multi-dimensions even with the use of super-computers.

To overcome this problem, we propose to use a Lagrangian technique
which exactly solves the transport stage on the entire domain and
then to project the solution on a grid to compute the contribution
of the collision operator. The resulting scheme shares many
analogies with semi-Lagrangian methods \cite{CrSon, CrSon1, Filbet}
and with Monte Carlo schemes \cite{sliu}, as we will explain, but on
the contrary to them, the method is as fast as a particle method
while the numerical solution remains fully deterministic, which
means that there is no source of statistical error. Thanks to this
approach we are able to compute the solution of the full six
dimensional kinetic equation on a laptop. This is, up to our
knowledge, the first time that the full kinetic equation has been
solved with a deterministic scheme on a single processor machine for
acceptable mesh sizes and in a reasonable amount of time (around ten
hours for $100^{3}$ space $\times$ $12^{3}$ velocity space mesh
points).

In this first work, we consider a simple collision operator,
\textit{i.e.} the BGK (Bhatnagar-Gross-Krook) relaxation operator
\cite{Gross}. The extension of the method to other operators like
the Boltzmann one \cite{bird, cercignani} or to other kinetic
equations like the Vlasov equation \cite{birsdall, Filbet} will be
considered in future works. At the present moment, the method is
designed to work on uniform grids, although extensions to other
meshes are possible and will be also considered in the next future.

The article is organized as follows. In section \ref{sec_Boltzmann},
we introduce the Boltzmann-BGK equations and their properties. In
section \ref{sec_DVM}, we present the discrete velocity model (DVM).
Then in section \ref{sec_num_approx} we present the numerical
scheme. Section \ref{sec:MonteCarlo} is devoted to the illustration
of the analogies between such fast kinetic scheme (FKS) and particle
methods. Several test problems up to three dimensional test cases
which demonstrate the capabilities and the strong efficiency of the
method are presented and discussed in section \ref{sec_tests}. Some
final considerations and future developments are finally drawn in
the last section.


\section{Boltzmann-BGK Equation}
\label{sec_Boltzmann}

We consider the following kinetic equation as a prototype model for
developing our method: \be
\partial_t f + v\cdot\nabla_{x}f = \frac1{\tau} (M_{f}-f),
\label{eq:B}
\ee
with the initial condition
\be
 f(x,v,t=0)=f_{0}(x,v).
\label{eq:B1} \ee This is the Boltzmann-BGK equation where
$f=f(x,v,t)$ is a non negative function describing the time
evolution of the distribution of particles which move with velocity
$v \in \R^d $ in the position $x \in \Omega \subset \R^{d}$ at time
$ t > 0$. For simplicity we consider the same dimension in space and
in velocity space $d$, however it is possible to consider different
dimensions in order to obtain different simplified models. In the
BGK equation the collisions are modeled by a relaxation towards the
local thermodynamical equilibrium defined by the Maxwellian
distribution function $M_{f}$. The local Maxwellian function is
defined by \be
 M_f=M_{f}[\rho,u,T](v)=\frac{\rho}{(2\pi \theta)^{d/2}}\exp\left(\frac{-|u-v|^{2}}{2\theta}\right) ,
\label{eq:M} \ee where $\rho \in \R^*$ and $u \in \R^d$ are the
density and mean velocity while $\theta=RT$ with $T$ the temperature
of the gas and $R$ the gas constant. The macroscopic values
$\rho$,$u$ and $T$ are related to $f$ by: \be \rho=\int_{\R^d} fdv,
\qquad u=\int_{\R^d} vfdv, \qquad  \theta=\frac{1}{\rho d}
\int_{\R^d}|v-u|^{2}fdv. \label{eq:Mo} \ee The energy $E$ is defined
by \be E=\frac{1}{2} \int_{\R^d}|v|^{2}fdv = \frac{1}{2} \rho |u|^2
+ \frac{d}{2} \rho \theta, \label{eq:E} \ee while the kinetic
entropy of $f$ by \be H(f)=\int_{\R^{d}}f\log f dv.
 \label{eq:H}
\ee The parameter $\tau> 0$ in (\ref{eq:B}) is the relaxation time.
In this paper, $\tau$ is fixed at the beginning of each numerical
test. Considering relaxation frequencies as functions of the
macroscopic quantities does not change the numerical scheme we will
propose and its behaviors. We refer to section \ref{sec_tests} for
the numerical values chosen.

If we consider the BGK equation (\ref{eq:B}) multiplied by $1$, $v$,
$\frac{1}{2}|v^{2}|$ (the so-called collision invariants), and then
integrated with respect to $v$, we obtain the following balance
laws: \be \ba \frac{\partial \rho}{\partial t} + \nabla_{x}\cdot
(\rho u) = 0, \ca \frac{\partial \rho u}{\partial t} +
\nabla_{x}\cdot (\rho u\otimes u+P) = 0, \ca \frac{\partial
E}{\partial t} +
 \nabla_x \cdot (Eu + Pu + q)  =  0,
\ea \label{eq:sys} \ee which express the conservation of mass,
momentum and total energy, in which $P=\int_{\R^{d}} (v-u)\otimes(v-u)f\, dv$
is the pressure tensor while $q=\int_{\R^{d}}\frac{1}{2}(v-u)|v-u|^2\, dv$ is
the heat flux. Furthermore the following inequality expresses the
dissipation of entropy: \be
\partial_{t} \left( \int_{\R^{d}} f\log f \, dv \right) + \nabla_{x} \cdot \left( \int_{\R^{d}} v f \log f \, dv \right) \leq 0 .
 \label{eq:HDI}
 \ee
System (\ref{eq:sys}) is not closed, since it involves other moments
of the distribution function than just $\rho$, $\rho u$ and $E$. Let
us describe one way to close the system.

The Maxwellian $M_{f}$ can be characterized as the unique solution
of the following entropy minimization problem \be H(M_{f})=
\min\left\{ H(f),\ f\geq 0 \ s.t. \int_{\R^{d}} m f \, dv =
U\right\} \label{eq:H1} \ee where $m$ and $U$ are the vectors of the
collision invariants and of the first three moments of $f$
respectively: \be m(v)= \left(1,v,\frac{1}{2}|v|^{2} \right), \
U=(\rho,\rho u,E). \label{eq:coll} \ee This is the well-known local
Gibbs principle, and it expresses that the local thermodynamical
equilibrium state minimizes the entropy, in the mathematical sense,
of all the possible states subject to the constraint that moments
$U$ are prescribed.

Formally, when the number of collision goes to infinity, which means
$\tau \rightarrow 0$, the function $f$ converges towards the
Maxwellian distribution. In this limit, it is possible to compute
the moments $P$ and $q$ of $f$ in terms of $\rho$, $\rho u$ and $E$.
In this way, one can close the system of balance laws (\ref{eq:sys})
and get the so-called Euler system of compressible gas dynamics
equations \be \ba \frac{\partial \rho}{\partial t} + \nabla_{x}
\cdot(\rho u) = 0, \ca \frac{\partial \rho u}{\partial t} +
\nabla_{x} \cdot (\rho u \otimes u+pI) = 0, \ca \frac{\partial
E}{\partial t} +\nabla_{x} \cdot((E+p)u) = 0, \ca p=\rho \theta,
\quad E=\frac{d}{2}\rho \theta +\frac{1}{2} \rho |u|^{2}. \ea
\label{eq:sys1} \ee


\section{The Discrete Velocity Model (DVM)}
\label{sec_DVM} The principle of Discrete Velocity Model (DVM) is to
set a grid in the velocity space and to transform the kinetic
equation in a set of linear hyperbolic equations with source terms.
We refer to the work of Mieussens \cite{Mieussens} for the
description of this model and we remind to it for the details.

Let $\mathcal{K}$ be a set of $N$ multi-indices of $\mathbb{N}^{d}$,
defined by $\mathcal{K}=\left\{k=(k^{(i)})_{i=1}^{d},\ k^{(i)}\leq
K^{(i)}\right\}$, where $\{K^{(i)}\}$ are some given bounds. We
introduce a Cartesian grid $\mathcal{V}$ of $\mathbb{R}^{d}$ by \be
\mathcal{V}=\left\{ v_{k}=k\Delta v+a, \ k \in
\mathcal{K}\right\},\ee where $a$ is an arbitrary vector of
$\mathbb{R}^{d}$ and $\Delta v$ is a scalar which represents the
grid step in the velocity space. We denote the discrete collision
invariants on $\mathcal{V}$ by
$m_{k}=(1,v_{k},\frac{1}{2}|v_{k}|^{2})$.

Now, in this setting, the continuous distribution function $f$ is
replaced by a $N-$vector $f_{\mathcal{K}}(x,t)$, where each
component is assumed to be an approximation of the distribution
function $f$ at location $v_{k}$: \be
f_{\mathcal{K}}(x,t)=(f_{k}(x,t))_{k},\qquad f_{k}(x,t) \approx
f(x,v_{k},t). \ee The fluid quantities are then obtained from
$f_{k}$ thanks to discrete summations on $\mathcal{V}$: \be
U(x,t)=\sum_{k}m_{k}f_k(x,t)\, \Delta v . \label{eq:DM} \ee The
discrete velocity BGK model consists of a set of $N$ evolution
equations for $f_k$ of the form \be
\partial_t f_{k} + v_{k} \cdot\nabla_{x}f_{k} = \frac1{\tau} (\E_{k}[U]-f_{k}),
\label{eq:DM1} \ee where $\E_{k}[U]$ is a suitable approximation of
$M_{f}$ defined next. Two strongly connected and important questions
arise when dealing with discrete velocity models. The first one is
about the truncation and boundedness of the velocity space. The
second one concerns the conservation of macroscopic quantities.

\paragraph{Truncation and boundedness of the velocity space.}
In DVM methods one needs to truncate the velocity space and to fix
some bounds. This gives the number $N$ of evolution equations
(\ref{eq:DM1}). Of course, the number $N$ is chosen as a compromise
between the desired precision in the discretization of the velocity
space and the computational cost, while the bounds are chosen to
give a correct representation of the flow. Observe in fact that, the
macroscopic velocity and temperature are bounded above by velocity
bounds. This implies that the discrete velocity set must be large
enough to take into account large variations of the macroscopic
quantities which may appear as a result of the time evolution of the
equations. Moreover, as a consequence of the velocity
discretization, we have that the temperature is bounded from below.
We summarize the above remarks by the following statement. Let $f$
be a non negative distribution function, then the macroscopic
velocity and temperature associated to $f$ in $\mathcal{V}$ by \be
u=\frac{1}{\rho}\langle vf\rangle_{\mathcal{K}}, \qquad
T=\frac{1}{dR\rho}\langle|v-u|^{2}f\rangle_{\mathcal{K}}, \ee where
$\langle .\rangle_{\mathcal{K}}$ denotes the summation over the set
of multi-indices $\mathcal{K}$, satisfy the bounds \cite{Mieussens}
\bea \min_{\mathcal{K}} v_{k}^{(i)}  \leq
& u^{(i)} & \leq \max_{\mathcal{K}}v_{k}^{(i)}, \; \; \forall i=1,\ldots,d \qquad  \\
\frac{1}{dR}\min_{\mathcal{K}}|v-u|^{2}\leq
&  T     & \leq \frac{1}{dR}\max_{\mathcal{K}}|v-u|^{2} .
\eea

\paragraph{Conservation of macroscopic quantities.}
Exact conservation of macroscopic quantities is impossible, because
in general the support of the distribution function is non compact.
Thus, in order to conserve macroscopic variables, different
strategies can be adopted, two possibilities are described in
\cite{gamba, Mieussens}. Moreover, the approximation of the
equilibrium distribution $M_f$ with $\E_{k}[U]$ must be carefully
chosen in order to satisfy the conservation of mass, momentum and
energy. In the following
section we will discuss our choices in details.
Such choices
prevent the lack of conservation of physical
quantities.
\begin{remark}{~}
Once DVM model is defined as above, the common choice which permits
to solve the kinetic equation is to discretize the $N$ evolution
equations with the preferred finite volume or finite difference
method \cite{Mieussens, Pal, Pal1, Pieraccini}. Alternatively, one
can reconstruct the distribution function in space and then follows
the characteristics backward in time to obtain the solution of the
linear transport equation \cite{CrSon, CrSon1, Filbet2, Filbet}. Our
choice, described in the next section, which enables to drastically
decrease the computational cost, consists of an exact solution of
the linear transport equation avoiding the reconstruction of the
distribution function.
\end{remark}

\section{Fast kinetic schemes (FKS)}
\label{sec_num_approx} The main features of the method proposed in
this work can be summarized as follows:
\begin{itemize}
\item The BGK equation is discretized in velocity space by using the DVM method.
\item A time splitting procedure is employed between the transport
and the relaxation operators for each of the resulting $N$ evolution
equations (\ref{eq:DM1}). First- and second-order Strang time
splitting \cite{Strang} are considered.
\item The transport part is exactly solved, which means without
using a spatial mesh. The initial data of this step is given by the
solution of the relaxation operator.
\item The relaxation part is solved on the grid. The initial data
for this step is given by the value of the distribution function in
the center of the cells after the transport step.
\end{itemize}
Before describing the scheme, we explain how we overcome the
drawback of the lack of conservation of macroscopic quantities in
DVM methods.

\subsection{Conservative methods} We introduce the conservative method for the initial
data and then we extend it to the scheme. The initialisation is done
in two steps. First we fix \be
\widetilde{f}_{k}(x,t=0)=f(x,v_k,t=0), \ k=1,\ldots,N.\ee Observe
that, in order to do this operation we do not need to discretize the
physical space, in others words, if the initial data is known
continuously, this information can be kept. However, for simplicity,
we already at this stage introduce a Cartesian uniform grid in the
physical space. This is defined by the set $\mathcal{J}$ of $M$
multi-indices of $\mathbb{N}^{d}$, which is
$\mathcal{J}=\{j=(j^{(i)})_{i=1}^{d},\ j^{(i)}\leq J^{(i)}\}$, where
$\{J^{(i)}\}$ are some given bounds which represent the boundary
points in the physical space. Next, the grid $\mathcal{X}$ of
$\mathbb{R}^{d}$ is given by \be \mathcal{X}=\{ x_{j}=j\Delta x+b, \
j \in \mathcal{J}\},\ee where $d$ represents at the same time the
dimension of the physical space and the dimension of the velocity
space which are taken equal for simplicity, even if this is not
necessary for the setting of the numerical method. Finally, $b$ is a
vector of $\mathbb{R}^{d}$ which determines the form of the domain
and $\Delta x$ is a scalar which represents the grid step in the
physical space. We consider a third discretization which is the time
discretization $t^{n}=n\Delta t$. We will later in the paper
introduce the time step limitations.

We denote with $f^{n}_{j,k}$ the approximation 
$f^{n}_{j,k}\simeq f(x_{j},v_{k},t_{n})$ and with
$\widetilde{f}^{n}_{j,k}$ the pointwise distribution value
$\widetilde{f}^{n}_{j,k}=f(x_{j},v_{k},t_{n})$ which are different,
for conservation reasons, as explained next. In this notation, the
discrete moments of the distribution $f$ are \be U_{j}^{n}=\langle
m_{k}f_{j,k}^{n}\, \Delta v\rangle_{\mathcal{K}}.\ee The
corresponding discrete equilibrium is denoted $\E_k[U_{j}^{n}]$, or
equivalently by $\E^{n}_{j,k}[U]$, which is an approximation of
$M_{f}[U_{j}^{n}]$ and it will be also defined later. When the
distribution function is truncated in velocity space, conservation
of the macroscopic quantities is no longer possible. Thus, in order
to restore the correct conserved variables we make use of a simple
constrained Lagrange multiplier method \cite{gamba}, where the
constraints are mass, momentum and energy of the solution.
Let us recall the technique from \cite{gamba}:
Let $N$
be the total number of discretization points of the velocity space
of the distribution function. We consider one space cell, the same
renormalization of $f$ should be considered for all spatial cells.
Let \be \widetilde{f} = \left(\widetilde{f}_1,
\widetilde{f}_2,\ldots,\widetilde{f}_N \right)^{T} \ee be the
pointwise distribution vector at $t=0$ and \be f = \left(f_1,
f_2,\ldots, f_N\right)^{T} \ee be the unknown corrected distribution
vector which fulfills the conservation of moments. Let \be
C_{(d+2)\times N}=\left(
\begin{array}{ll}
& (\Delta v)^{d}\\
& v_k(\Delta v)^{d} \\
& |v_{k}|^{2}(\Delta v)^{d}\\
\end{array}
\right) \ee and $U_{(d+2)\times 1} = (\rho \ \rho u \ E)^{T}$ be the
vector of conserved quantities. Conservation can be imposed using a
constrained optimization formulation:
\begin{eqnarray}
  &\nonumber  \mbox{ Given } \widetilde{f}\in \mathbb{R}^{N}, \ C \in \mathbb{R}^{(d+2)\times
  N}, \mbox{ and } U \in\mathbb{R}^{(d+2)\times 1},\label{eq:minim}\\
  & \mbox{ find } f \in \mathbb{R}^{N} \mbox{ such that } \\
  &\nonumber \|\widetilde{f} -f\|^{2}_{2} \mbox{ is minimized subject to the constrain } Cf = U.
\end{eqnarray}

To solve this constrain minimization problem, one possibility is to
employ the Lagrange multiplier method. Let $\lambda\in
\mathbb{R}^{d+2}$ be the Lagrange multiplier vector. Then the
corresponding scalar objective function to be optimized is given by
\be L(f, \lambda) = \sum_{k=1}^{N} |\widetilde{f}_{k}-f_{k} |^{2} +
\lambda^{T} (Cf- U) . \ee The above equation can be solved
explicitly. In fact, taking the derivative of $L(f, \lambda)$ with
respect to $f_{k}$, for all $k = 1, ...,N$ and $\lambda_i$, for all
$i = 1, ..., d + 2$, that is to say the gradient of $L$, we obtain
\be \frac{\partial L}{\partial f_{k}} = 0, \; \; k = 1, ...,N \
\Longrightarrow \ f = \widetilde{f} + \frac{1}{2} C^{T}\lambda,
\label{eq:legrange} \ee and \be \frac{\partial L}{\partial
\lambda_{i}} = 0, \; \; i = 1, ...,d+2 \ \Longrightarrow \ Cf = U.
\label{eq:legrange2} \ee Now, solving for $\lambda$ we get \be
CC^{T}\lambda = 2(U- C\widetilde{f}),\ee and observing that the
matrix $CC^{T}$ is symmetric and positive definite, since $C$ is the
integration matrix, one deduces that the inverse of $CC^{T}$ exists.
In particular the value of $\lambda$ is uniquely determined by \be
\lambda= 2(CC^{T})^{-1}(U-C \widetilde{f}) . \ee Back substituting
$\lambda$ into (\ref{eq:legrange}) provides \be f = \widetilde{f} +
C^{T} (CC^{T})^{-1}(U-C\widetilde{f}). \label{eq:minim1} \ee Observe
that, following the same principle, we can impose the conservation
of other macroscopic quantities, in addition to mass, momentum and
energy. A key point is that, in practice, we need to solve the above
minimization problem only for the initial data $f(x_{j},v_{k},t=0)$
because once the conservation is guaranteed for $t=0$, this is
also guaranteed for the entire computation because the exact
solution is used for solving the transport step. The only possible
source of loss of conservation for the entire scheme is the
relaxation step. This means that, for this step, we will need to
impose conservation of the macroscopic quantities but only for the
equilibrium distribution.

The discretization of the Maxwellian distribution $M_f(x,v,t)$,
should satisfy the same properties of conservation of the
distribution $f$, \textit{i.e.} $ U_{j}^{n}=\langle
m_{k}f_{j,k}^{n}\, \Delta v\rangle_{\mathcal{K}}=\langle
m_{k}\E_{k}[U_{j}^{n}]\, \Delta v\rangle_{\mathcal{K}}$. To this
aim, observe that the natural approximation \be
\E_k[U^{n}_{j}]=M_{f}(x_{j},v_{k},t_{n}), \; k\in\mathcal{K}, \;
n\geq0, \; j\in\mathcal{J}\ee cannot satisfy these requirements, due
to the truncation of the velocity space and to the piecewise
constant approximation of the distribution function. Thus, the
calculation carried out above for the definition of the initial
distribution $f$, can be also performed for the equilibrium
distribution $M_f$. This should be done each time we invoke the
equilibrium distribution during the computation. The function
$\E[U]$ is therefore given by the solution of the same minimization
problem defined in (\ref{eq:minim}), and its explicit value is given
mimicking (\ref{eq:minim1})
 by \be \E[U] = M_f[U] + C^{T}
(CC^{T})^{-1}(U-CM_f[U]),\label{eq:minimMax}\ee where $M_f[U]$
represents the pointwise values of the Maxwellian distribution
$M_f[U]=M_{f}(x_{j},v_{k},t_{n})$. Notice that the computation of
the new distributions $f$ and $\E$ only involves a matrix-vector
multiplication. In fact, matrix $C$ only depends on the parameter of
the discretization and thus it is constant in time. In other words
matrices $C$ and $C^{T}(CC^{T})^{-1}$ can be precomputed and stored
in memory during the initialisation step. They are used during the
simulation when the solution of system (\ref{eq:minim}) is invoked.

Another possibility to approximate the Maxwellian distribution
$M_{f}$ is proposed in \cite{Mieussens}. In that work, the authors
define $\E_{k}[U]$ as the solution of a discrete entropy
minimization problem \be H_{\mathcal{K}}(\E[U])
=\min\left\{H_{\mathcal{K}}(g), g\geq 0\in\mathbb{R}^{N} \mbox{ such
that } \langle mg\rangle_{\mathcal{K}}=U\right\}. \label{eq:defE}
\ee This discretization (existence, uniqueness, convergence) has
been mathematically studied in~\cite{Mieussens}. However, one
drawback of this method, is the need for solving a non
linear system of equations in each spatial cell for each time step.
As we seek for efficiency, we only consider the first minimization
strategy (\ref{eq:minim}) to approximate the equilibrium
distribution $M_f$.

\subsection{Conservative and fast kinetic schemes FKS}
We can now present the full scheme. Let us start with the
first-order splitting scheme and then, define the second-order in
time method based on the Strang splitting strategy \cite{Strang}.

Let $f^{0}_{j,k}$ be the initial data defined as a piecewise
constant function in space and in velocity space, solution of
equation (\ref{eq:minim1}) with
$\widetilde{f}^{0}_{j,k}=f(x_{j},v_{k},t=0)$. We recall that the
choice of a piecewise constant function in space is not mandatory
for the method. Let also $\E^{0}_{j,k}[U]$ be the initial
equilibrium distribution solution of equation (\ref{eq:minimMax})
with $M^{0}_{j,k}=M_f(x_{j},v_{k},t=0)$. We start describing the
first time step of the method $[t^0;t^1]$ starting at $t^0=0$, we
further generalize the method to the generic time step
$[t^n;t^{n+1}]$ starting from $t^n$.

\paragraph{First time step $[t^0;t^1]$.} Let us describe the transport
and relaxation stages. \\
\textit{Transport stage.} We need to solve $N$ linear transport
equations of the form: \be
\partial_t f_{k} + v_{k} \cdot\nabla_{x}f_{k}=0, \quad k=1,\ldots,N
\label{eq:DVM}, \ee where the initial data for each of the $N$
equations is a piecewise constant function in the three dimensional
space defined as \be \overline{f}_{k}(x,t^0=0)=f^{0}_{j,k} \quad
\forall x \in [x_{j-1/2},x_{j+1/2}], \quad  k=1,\ldots,N. \ee The
exact solution of the $N$ equations at time $t^1=t^0+\Delta t =
\Delta t$ is given by \be
\overline{f}_{k}(x,t^1)=\overline{f}^{*}_{k}(x)
                           =f(x-v_{k}\Delta t), \quad k=1,\ldots,N.
\ee Observe that, here, we do not need to reconstruct our function
as for instance in the semi-Lagrangian schemes \cite{Filbet2,
Filbet}, the shape of the function in space is in fact known and
fixed at the beginning of the computation. Once the solution of the
transport step is known, to complete one step in time, we need to
compute the solution of the relaxation step. As in finite volume or
finite difference methods, we solve the relaxation step only on the
grid, thus only the value of the distribution function $f$ in the
centers of the cells are computed. From the exact solution of the
function $f_{k}$ we can immediately recover these values at the cost
of one simple vector multiplication. On the other hand, one notices
that for classical finite difference or finite volume methods nested
loops for each dimension in space and in velocity space are
mandatory to compute the solution of the transport part. This makes
the computational cost of these methods extremely demanding in the
multidimensional cases. On the contrary, the computational cost of
the method we propose is only of the order of the number of points
in which the velocity space is discretized ($O(N)$). In particular,
for uniform meshes, we only need to compute the new value of $f_k$
in the center of one single cell, to know the solution in the center
of all others cells.

\textit{Relaxation stage.} For this step we need to locally solve on
the grid, \textit{i.e.} in the center of each spatial cell, an
ordinary differential equation. Thus, we have to solve: \be
\partial_t f_{j,k} =\frac{1}{\tau}(\E_{j,k}[U]-f_{j,k}), \ \ k=1,\ldots,N, \ \ j=1,\ldots,M
\label{eq:relax},
\ee
where the initial data is the result of the
transport step
\be \overline{f}_{k}(x_{j},t^1) \equiv
f^{*}_{k}(x_{j}), \ \ k=1,\ldots,N , \ \ j=1,\ldots,M.
\ee
Any
discretization method in time for this term can be chosen, as for instance
the preferred Runge Kutta method. However, being the above equation
a first order linear ordinary differential equation, we choose to
compute the exact solution. The last ingredient needed to perform the
computation, is the value of the equilibrium distribution $\E$ at
the center of the cell after the transport stage. To this aim,
observe that, the Maxwellian distribution does not change during the
relaxation step, which means that during this step the macroscopic
quantities remain constants. This implies that only the transport
stage possibly modifies the equilibrium distribution. In order to
compute the Maxwellian, the macroscopic quantities in the center of
the cells, \textit{i.e.} the density, the mean velocity and the
temperature, are given by summing the local value of the discrete
distribution $f$ over the velocity set $\langle
m_{k}f^{*}_{j,k}\Delta v\rangle_{\mathcal{K}}=U^{1}_{j}, \
j=1,\ldots,M$, where $f^{*}_{j,k}=f^{*}_{k}(x_{j})$. Finally, the
discrete equilibrium distribution at time $t^1=t^0+\Delta t$ is the
solution of equation (\ref{eq:minimMax}) with moments $U^{1}_{j}, \
j=1,\ldots,M$. We can now compute the solution of the relaxation
stage by \be f^{1}_{j,k}= \exp(-\Delta
t/\varepsilon)\overline{f}^{*}_{j,k}+(1-\exp(-\Delta
t/\varepsilon))\E^{1}_{j,k}[U]. \ee Observe that the above equation
furnishes only the new value of the distribution at time
$t^1=t^0+\Delta t=\Delta t$ in the center of each spatial cell for
each velocity $v_k$. However, what we need, in order to continue the
computation, is the value of the distribution $f$ in all points of
the space. To overcome this problem, in classical discrete velocity
methods several authors \cite{Mieussens, Pal} consider the
distribution function constant in the cell as well as the Maxwellian
distribution. The result is that they need to solve only an ordinary
differential equation in the center of the cell taking the average
value of the macroscopic quantities inside one cell. Here, we make a
different approximation. We consider that the equilibrium
distribution $M_f$ has the same form as the distribution $f$ in
space. In other words $\E_k$ is a piecewise constant function in
space for each velocity $v_k$. The values of this piecewise constant
function are the values computed in the center of the spatial cells,
\textit{i.e.} one defines
\be \overline{\E}_{k}(x,t^1)=\E^{1}_{j,k},
\; \; \forall x \; \mbox{s.t.} \; \;
\overline{f}_{k}(x,t^1)=\overline{f}_{k}(x_j,t^1), \ j=1,\ldots,M.
\ee
This further implies that the relaxation term writes in term of
spacial continuous function $\overline{f}_{k}(x,t^1)$ as \be
\overline{f}_{k}(x,\Delta t) =
  \exp(-\Delta t/\varepsilon)\overline{f}_{k}(x,t^1)
  +(1-\exp(-\Delta t/\varepsilon))\overline{\E}_{k}(x,t^1)[U].
\ee For each velocity $v_k$ this choice permits to keep the
form of the distribution $f_k$ constant in space throughout the computation,
and, as a consequence it drastically reduces the computational cost.
This ends the first time step. \\

We focus now on the time marching procedure for the first- and
second-order splitting schemes which will allow to solve the
Bolzmann-BGK equation.

\paragraph{Generic time step $[t^n;t^{n+1}]$.} We present
a first-order and second-order Strang splitting technique
\cite{Strang}.

\textit{First-order splitting:} Given the value of the distribution
function $\overline{f}^{n}_{k}(x)$, for all $k=1,\ldots,N$, and all
$x \in \mathbb{R}^{d}$ at time $t^n$, the value of the distribution
at time $t^{n+1}$, $\overline{f}_{k}^{n+1}(x)$, is given by \be
\overline{f}_{k}^{*}(x)=f^{n}_{k}(x-v_k\Delta t), \quad k=1,\ldots,N
\ee \be \overline{f}^{n+1}_{k}(x)=\exp(-\Delta
t/\varepsilon)\overline{f}^{*}_{k}(x)+(1-\exp(-\Delta
t/\varepsilon))\overline{\E}^{n+1}_{k}(x)[U], \quad k=1,\ldots,N,
\label{relaxdvm} \ee where $\overline{\E}^{n+1}_{k}(x)[U]$ is a
piecewise constant function, computed considering the solution of
the minimization problem (\ref{eq:minimMax}) relative to the moments
value in the center of each spatial cell after the transport stage:
$U^{n+1}_{j}, \ j=1,\ldots,M$. These moments are given by computing
$\langle m_{k}f^{*}_{j,k}\Delta v\rangle_{\mathcal{K}}$ where
$f^{*}_{j,k}$ is the value that the distribution function takes
after the transport stage in the center of each spatial cell.

\textit{Second-order splitting:}
Given the value of the distribution function
$\overline{f}^{n}_{k}(x), \ k=1,\ldots,N,$ $x \in \mathbb{R}^{d}$ at
time $t^n$, the scheme reads \be
\overline{f}_{k}^{*}(x)=f^{n}_{k}(x-v_k\Delta t/2), \quad k=1,\ldots,N
\ee \be \overline{f}^{**}_{k}(x)=\exp(-\Delta
t/\varepsilon)\overline{f}^{*}_{k}(x)+(1-\exp(-\Delta
t/\varepsilon))\overline{\E}^{*}_{k}(x)[U], \quad k=1,\ldots,N, \ee
where $\overline{\E}^{*}_{k}(x)[U]$ is a piecewise constant
function, computed considering the solution of the minimization
problem (\ref{eq:minimMax}) relative to the moments values in the
center of each spatial cell after the transport stage of size
$\Delta t/2$. We call these moments $U^{*}_{j}, \ j=1,\ldots,M$. They are
given by the discrete summation $\langle m_{k}f^{*}_{j,k}\Delta
v\rangle_{\mathcal{K}}$ where $f^{*}_{j,k}$ is the value that the
distribution function takes after the transport stage in the center
of each spatial cell. The last step consists of a second transport
stage of half time step \be
\overline{f}_{k}^{n+1}(x)=f^{**}_{k}(x-v_k\Delta t/2), \quad
k=1,\ldots,N, \ee which ends the second-order splitting scheme.

\begin{remark}{~}
\begin{itemize}
\item As already mentioned the choices of uniform meshes and piecewise
constant functions in space are not necessary for the construction
of the method. These choices have been made because we wanted to
analyze the method in its simplest form. We postpone to future works
the study of non-uniform meshes and different shapes of the
distribution function $f$ in space. However, a key point is that,
even if the method in its general form is already much more faster
than finite volume, finite difference or semi Lagrangian methods for
kinetic equations, it can be made extremely fast in the case of
uniform meshes as we will explain in the next paragraph.

  \item For finite volume or finite difference
methods applied to discrete velocity models of kinetic equations,
the second-order time splitting implies the computation of the
transport stage in two steps, from $t^n$ to $t^{n+1/2}$ and from
$t^{n+1/2}$ to $t^{n+1}$. Conversely the same operation can be done
with the relaxation step to get second order accuracy.
\item In our method, extending the scheme from first- to
second-order time splitting is almost as expensive as the first-order. In
fact, except for the first time step in which we need to compute two
times the transport operator with $\Delta t/2$, starting from the
second time step we have to solve a sequence of two $\Delta t/2$
transport stages. However, being the transport computed exactly,
solving the linear transport equations two times with $\Delta t/2$
or only one time with the entire $\Delta t$ provides the same solution.
This means that, in order to obtain second-order accuracy
it is sufficient to solve the first time step with $\Delta t/2$ and
then proceed as for the first-order method to obtain global
second-order accuracy in time.
\item However, any time splitting method does degenerate
to first-order accuracy in the fluid limit, that is to say, when
$\tau\rightarrow 0$.
\item Due to the fact that the
relaxation stage preserves the macroscopic quantities, the scheme is
globally conservative. In fact, at each time step, the change of
density, momentum and energy is only due to the transport step. This
latter, being exact, does preserve the macroscopic quantities as
well as the distribution function.
\item For the same reason, the scheme is
also unconditionally positive. In others words, we observe that
$f^{n}_{k}(x)\geq 0$, for all $n>0$, and $k=1,\ldots,M$ if the
initial datum is positive $f^{0}_{k}(x)\geq 0$ for all
$k=1,\ldots,M$. In fact, the transport maintains the shape of $f$
unchanged in space while the relaxation towards the Maxwellian
distribution is a convex combination of $M_f$ and $f(x-v_k\Delta t)$
both being positive.
\item We expect the scheme to perform very well in collisionless or
almost collisionless regimes. In these cases in fact the relaxation
stage is neglectible and only the exact transport does play a role.
When moving from rarefied to dense regimes the projection over the
equilibrium distribution becomes more important. Thus, the accuracy
of the scheme is expected to diminish in fluid regimes, because the
projection method is only first-order accurate. One possibility, for
such regimes is to increase the order of the projection method
towards the equilibrium. This possibility will also be analyzed in
future works.
\item The time step
$\Delta t$ is chosen as the classical CFL condition \be \Delta t
\max_{k} \left( \frac{|v_{k}|}{\Delta x} \right) < 1 .
\label{eq:Time} \ee Observe that this choice is not mandatory, in
fact the scheme is always stable for every choice of the time step,
but being based on a time splitting technique the error is of the
order of $\Delta t$ or $(\Delta t)^{2}$. This suggests to take the
usual CFL condition in order to maintain the error small enough.
\end{itemize}
\end{remark}

\section{Analogies with particles methods}
\label{sec:MonteCarlo} In this section we first introduce a Monte
Carlo particle method which permits to solve the Boltzmann-BGK
equation. Next, we introduce its deterministic counterpart,
\textit{i.e} a deterministic particle method. Finally, we show that
a slightly modified version of this latter method, in which the
positions of the particles, instead of being randomly chosen, are
taken initially at the same position in space for all the cells, is
equivalent to a FKS method where some specific choices of the
discretization parameters are done. This analogy permits to derive a
very convenient form of the algorithm which 
for this choice of the discretization parameters.

The starting point of Monte Carlo methods is again given by a time
splitting between free transport \be
\partial_t f + v\cdot\nabla_{x}f
=0,\ee and collision, which in the case of the BGK operator is
substituted by a relaxation towards the equilibrium \be
\partial_t f=\frac{1}{\tau}(f-M_f[U]).\label{eq:9}\ee
In Monte Carlo simulations the distribution function $f$ is
discretized by a finite set of particles
\begin{eqnarray}
& & f = \sum_{i=1}^N \mathfrak{m}_i \, \delta(x-x_i(t))
\delta(v-v_i(t)), \label{particle}
\end{eqnarray}
where $x_i(t)$ represents the particle position, $v_i(t)$ the
particle velocity and $\mathfrak{m}_i$ the particle mass which is
usually taken constant. During the transport stage the
particles move to their next positions according to \be x_i(t+\Delta
t)=x_i(t)+v_i(t)\Delta t, \label{transport}\ee where $\Delta t$ is
such that an appropriate CFL condition holds. This condition
normally implies that one particle does not cross more than one cell
in one time step.

The collision step acts only locally, changes the velocity
distribution but preserves the macroscopic quantities. In this case,
as already explained, the space homogeneous problem admits the
following exact solution at time $t+\Delta t$ \be f(t+\Delta
t)=e^{-\Delta t/\tau}f(t)+(1-e^{-\Delta
t/\tau})M_f[U](t).\label{eq:8}\ee Thus, in a Monte Carlo method, the
relaxation step consists in replacing randomly selected particles
with Maxwellian particles with probability $(1-e^{-\Delta t/\tau})$.
This means\be v_i(t+\Delta t)=\left\lbrace
\begin{array}{ll}
\displaystyle v_i(t), &  \text{with probability} \ e^{-\Delta
t/\tau}\\
\displaystyle M_f[U](v), &  \text{with probability} \ 1-e^{-\Delta
t/\tau}\\
\end{array},
\right.\label{relax}\ee where $M_f[U](v)$ in the above expression
represents a particle sampled from the Maxwellian distribution with
moments $U$. Observe that, second-order splitting can be used as
well in the Monte Carlo methods. As in the case of the FKS, because
the transport step is resolved exactly, the change with respect to
the first-order method is only the first time step which has to be
computed with a time step of $\Delta t/2$. This will assure
second-order accuracy in time except in the limit $\tau\rightarrow
0$ in which the method degenerates again to first-order accuracy.

We introduce now a modified particle method which shares many
analogies with our method. Instead of the continuous kinetic
equation, this modified particle approach solves the discrete
velocity approximation of the kinetic equation. In this method, the
distribution function is again represented by a piecewise constant
function, defined on a compact support in the velocity space. The
distribution function is approximated by a finite set of particles
in each spatial cell as in the previous Monte Carlo method. The main
difference with respect to the other particle method is that now the
particles can attain only a discrete set of velocities and that the
mass of each particle is no more a constant, instead it changes in
time during the time evolution of the kinetic equation. These types
of methods are known in literature as weighted particles methods
\cite{Mas2, Mas, Mas3}. Therefore we consider
\begin{eqnarray}
& & f = \sum_{i=1}^N \mathfrak{m}_i(t) \, \delta(x-x_i(t)), \quad
\delta(v-v_i(t)), \quad v_i(t)=v_k, \quad  k \in \mathcal{K},
\label{particle2}
\end{eqnarray}
where $\mathcal{K}$ is the same set of multi-indices than the DVM
discretization (this means that the number of particle is fixed
equal to the number of points $N$ in which the velocity space is
discretized). The BGK equation is again split into two stages: a
transport and a relaxation stage. The transport part, as before,
corresponds to the motion of the particles in space caused by their
velocities (\ref{transport}). The main difference is in the solution
of the relaxation part (\ref{eq:9}). In order to solve this equation
from a particle point of view, we change the mass of each particle
using the exact solution of the relaxation equation,  \textit{i.e.}
\be f(t+\Delta t)=e^{-\Delta t/\tau}f(t)+(1-e^{-\Delta
t/\tau})M_f[U](t).\label{eq:relax2}\ee this corresponds to \be
\mathfrak{m}_i(t+\Delta t)=e^{-\Delta t/\tau}f(v_i)+(1-e^{-\Delta
t/\tau})\E(v_i)[U], \quad  i=1,\ldots,N. \label{eq:9bis}\ee Again in
practice to avoid the loss of conservation of macroscopic
quantities, once the conserved quantities are computed in one cell,
we solve the minimization problem (\ref{eq:minimMax}) to get the
function $\E[U]$. Thus, the above procedure requires the knowledge
of $U_{j}, \ j=1,\ldots,M$, which can only be estimated from the
sample positions. The simplest method, which produces a piecewise
constant reconstruction, is based on evaluating the histogram of the
samples on the grid, considering all the samples inside one cell be
of the same importance irrespectively of their positions. In
practice, the density $\rho_{j}, \ j=1,\ldots,M$ is given by the
number of samples $N_{I_j}$ belonging to the cell $I_j$ \be
\rho_{j}=\frac{1}{\Delta x}\sum_{x_i\in I_j} \mathfrak{m}_i,
\label{rec} \ee while the mean velocity in each spatial direction
and the energy are given by \be u_{j}=\frac{1}{\rho_j}\sum_{x_i\in
I_j} \mathfrak{m}_iv_{i}, \quad \quad E_{j}=\frac{1}{2\Delta x}
\sum_{x_i\in I_j}\mathfrak{m}_i|v_{i}|^{2} . \label{rec1} \ee The
method described above deserves some remarks. First, note that as
$\tau\to 0$ the method becomes a particle scheme for the limiting
fluid dynamic equations. This limit method is the analogous of a
kinetic particle method for the compressible Euler equations.
Second, the simple splitting method described is first-order in
time. Second order Strang splitting can be implemented similarly to
the case of the FKS scheme described in the previous section.

Now, we dispose of all the elements which permit to highlight the
similarities with the FKS scheme. Observe that the relaxation step
(\ref{eq:9bis}) is no more solved statistically as for the original
Monte Carlo method (\ref{relax}). Thus, the scheme described is in
fact a deterministic particle scheme, in which, however, the
particle positions are still randomly initialized. Now, if we
consider the piecewise reconstruction of the macroscopic quantities
introduced before (\ref{rec}-\ref{rec1}), we take one single
particle for each velocity $v_k, k\in \mathcal{K}$ and we fix all
particles positions at the beginning of the computation at the
center of each cell we obtain the FKS described in the previous
section. In fact, first the number of particles in each spatial cell
remains constant in time and equal to the number of mesh point in
velocity space $N$. This is because for each particle that goes out
of one cell, there exists another particle with the same velocity
which enters in the cell from another location. This is due to the
fact that particles have initially the same position, they never
change velocity and the mesh is uniform. Thus, during the time
evolution the only quantity that is modified is the mass of the
particle. This mass changes according to the solution of the
relaxation equation (\ref{eq:9bis}). This is exactly what happens in
the FKS method in equation (\ref{relaxdvm}). Finally, the transport
is solved exactly for the particle scheme as well as for the FKS
method. However, the weighted particle scheme, can be viewed as a
particular case of the FKS method. In fact, to regain the weighted
particle method, we have to fix the position of the particles, take
only a single particle for a given velocity $v_k$, the mesh must be
uniform and the shape of the distribution function in space must be
piecewise constant for the FKS method. This analogy between the two
schemes permits, from one side, to derive a very efficient algorithm
for the FKS method. From the other side, it opens the way to in deep
discussions from the theoretical point of view on the relation
between the two methods , like the different convergence properties
of the two approaches. We remind to a future work for an analysis of
the convergence of the FKS method.

\section{Numerical tests}
\label{sec_tests}


\subsection{General setting}
\label{subsec_gen_set}
\begin{figure}[h!]
\begin{center}
\includegraphics[scale=0.43]{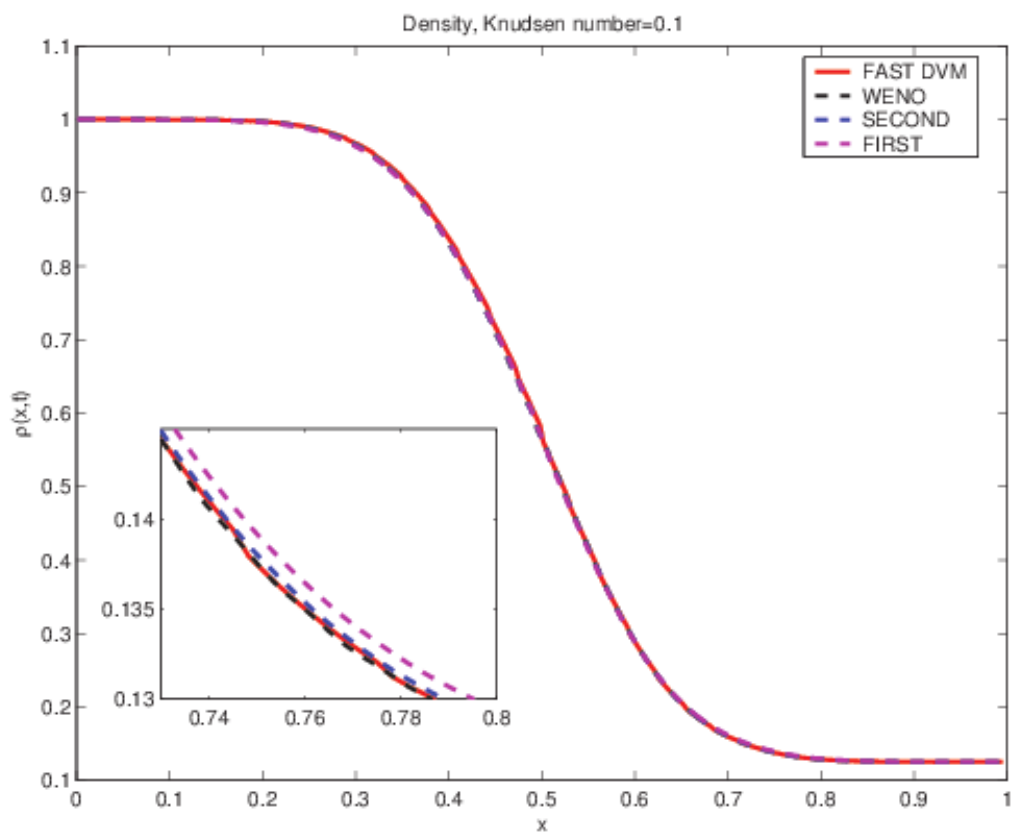}
\includegraphics[scale=0.43]{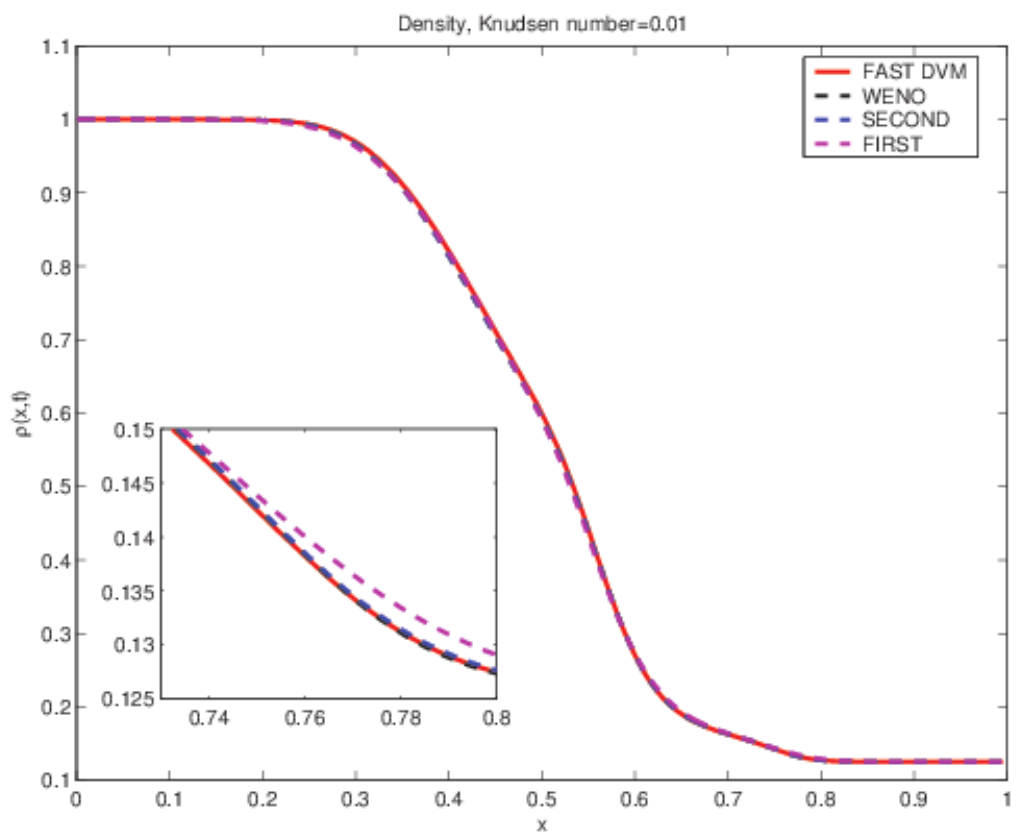}\\
\includegraphics[scale=0.43]{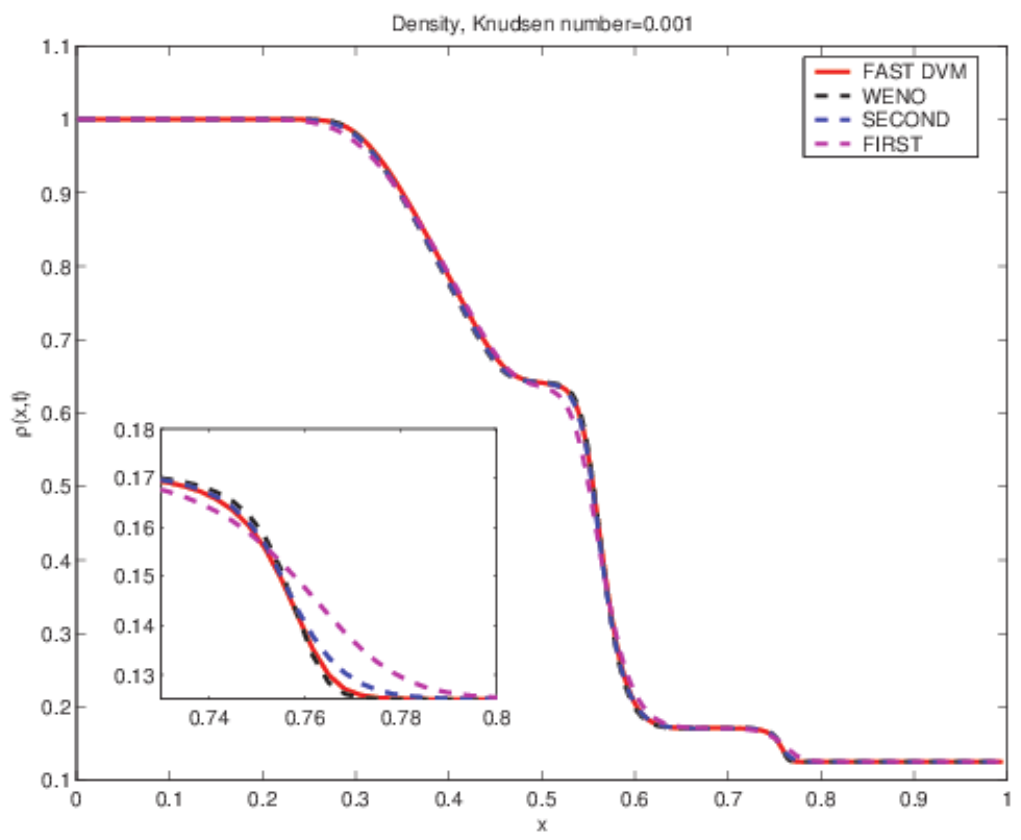}
\includegraphics[scale=0.43]{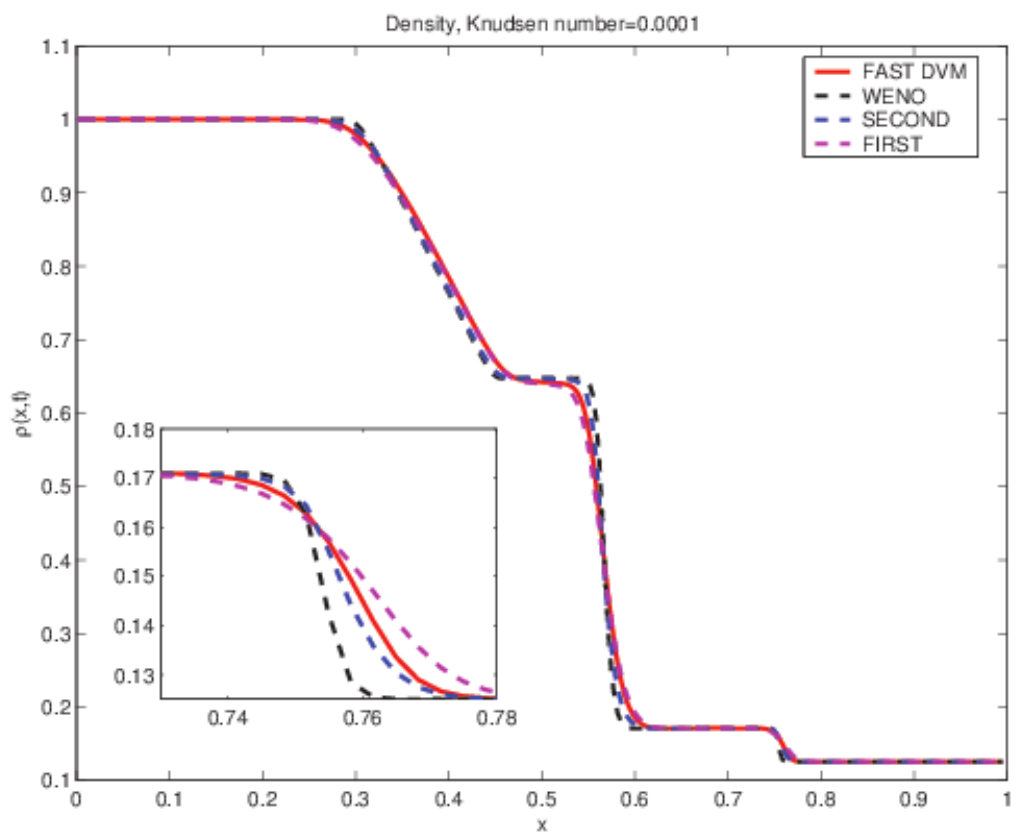}\\
\caption{Sod test: solution at $t_{\text{final}}=0.05$ for the
density, with $\tau=10^{-1}$ (top left), $\tau=10^{-2}$ (top right),
$\tau=10^{-3}$ (bottom left) and $\tau=10^{-4}$ (bottom right).}
\label{sod1}
\end{center}
\end{figure}

In this section, we present several numerical tests to illustrate
the main features of the method. First the performance of the scheme
is tested in the one dimensional case for solving the Sod problem.
In this case, we do comparisons of our method with different finite
difference methods which can solve the same problem. In the one
dimensional case, the computational speedup is not very relevant
being all classical methods sufficiently fast. However, the FKS
method is still faster than the other methods. In a second series of
tests we solve a two dimensional- two dimensional kinetic equation.
Finally we solve a full three-three dimensional problem. In this
situation, it is a matter of fact that computing the solution of a
kinetic equation with finite difference, finite volume or
semi-Lagrangian methods is unreasonable. We will show results from
our method running on a mono-processor laptop machine.


\subsection{1D Sod shock tube problem}
\label{subsec_sod}

\begin{figure}[h!]
\begin{center}
\includegraphics[scale=0.425]{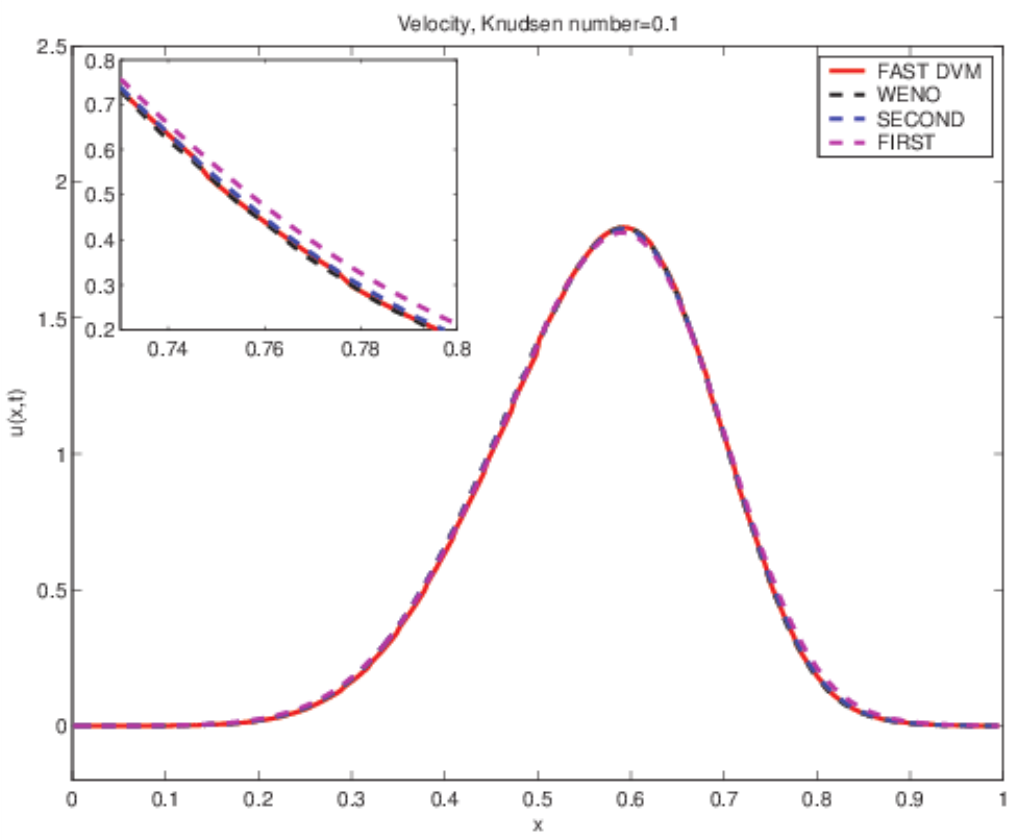}
\includegraphics[scale=0.425]{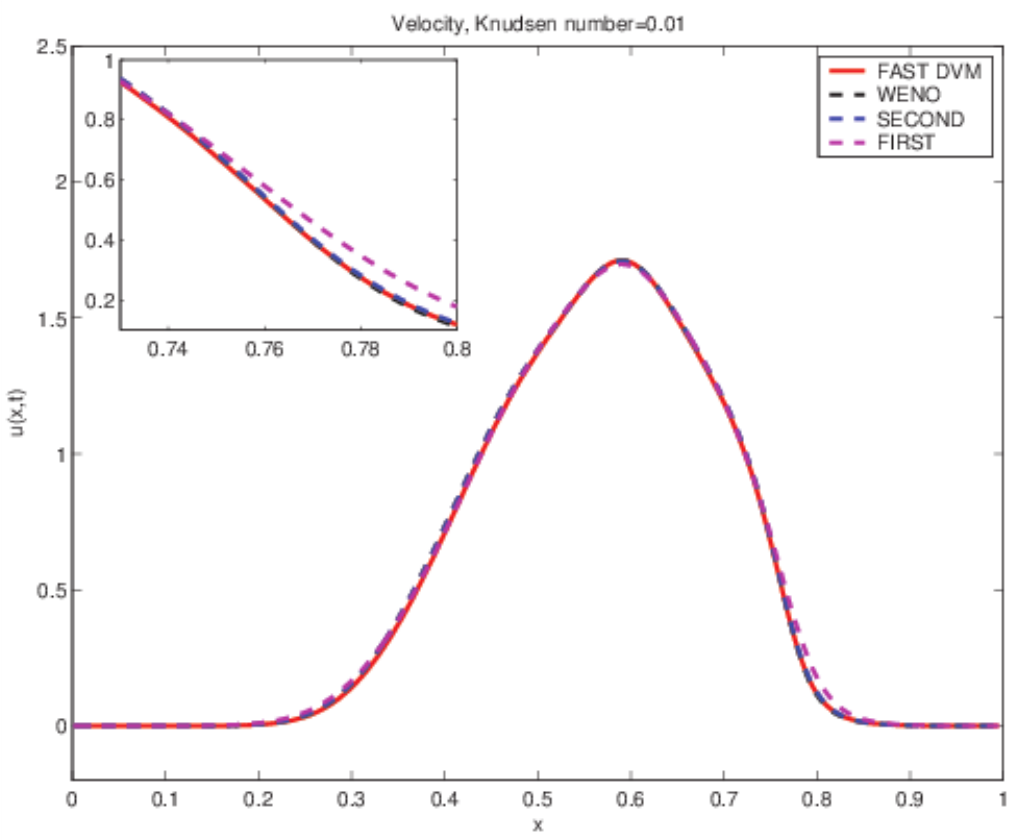}\\
\includegraphics[scale=0.425]{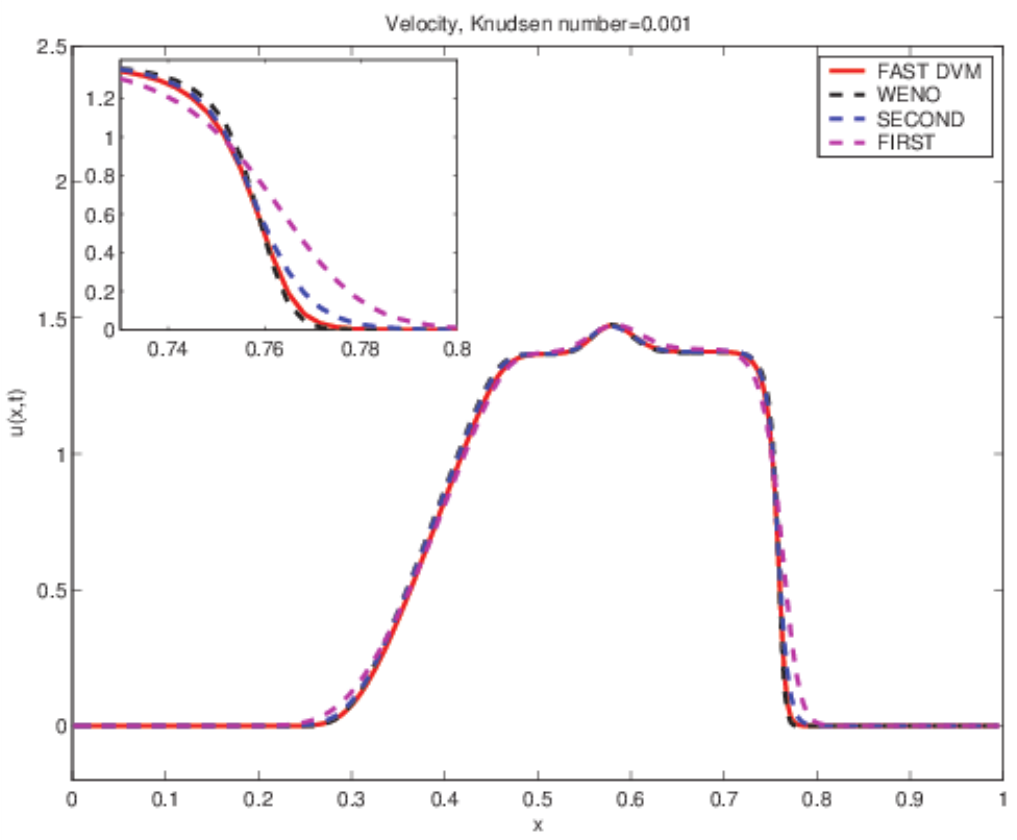}
\includegraphics[scale=0.425]{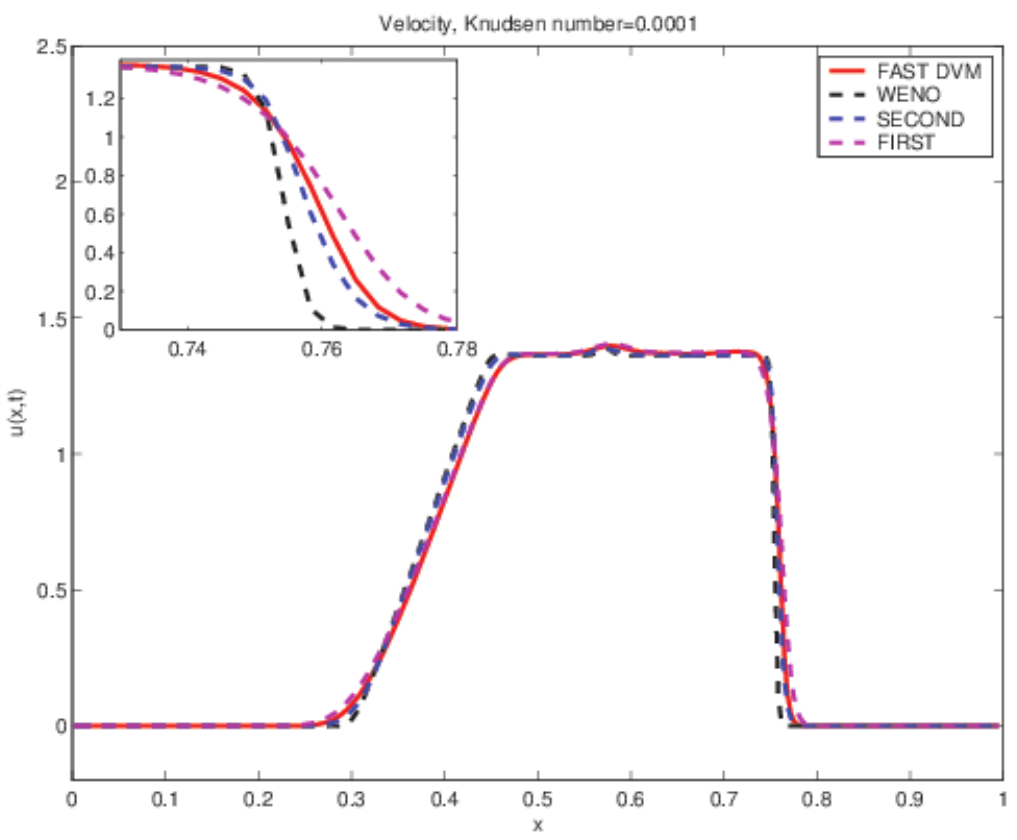}\\
\caption{1D Sod test: solution at $t_{\text{final}}=0.05$ for the
mean velocity, with $\tau=10^{-1}$ (top left), $\tau=10^{-2}$ (top
right), $\tau=10^{-3}$ (bottom left) and $\tau=10^{-4}$ (bottom
right).} \label{sod2}
\end{center}
\end{figure}

\begin{figure}[h!]
\begin{center}
\includegraphics[scale=0.43]{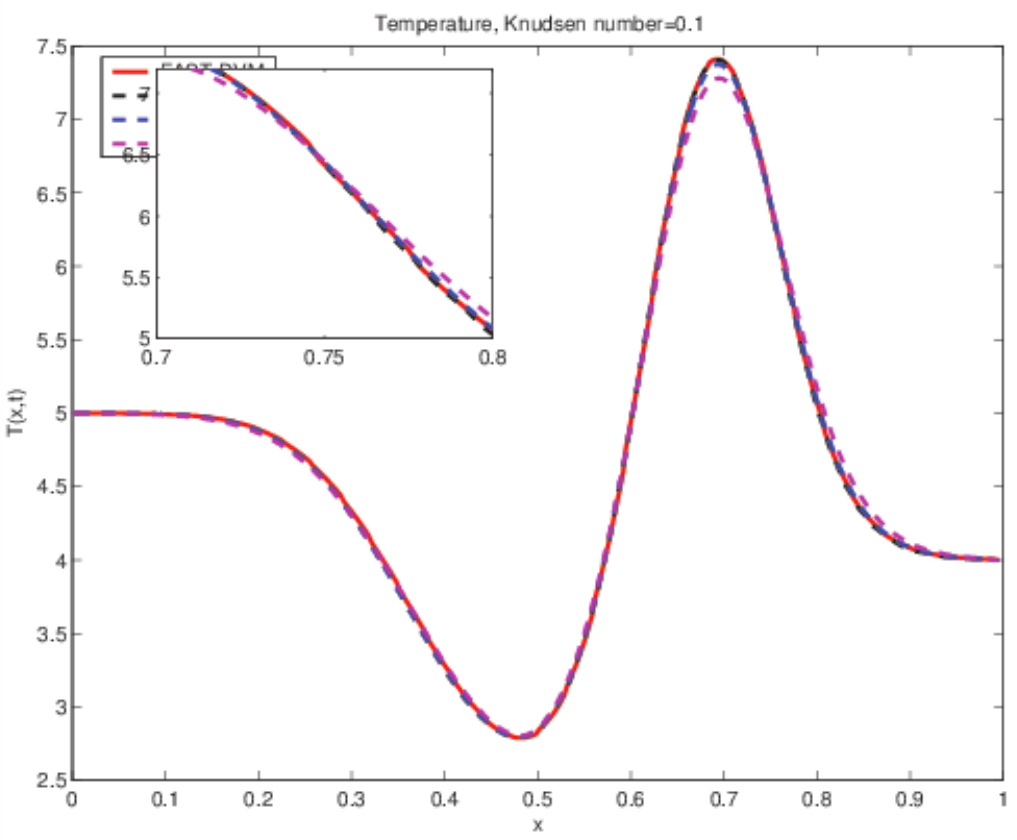}
\includegraphics[scale=0.43]{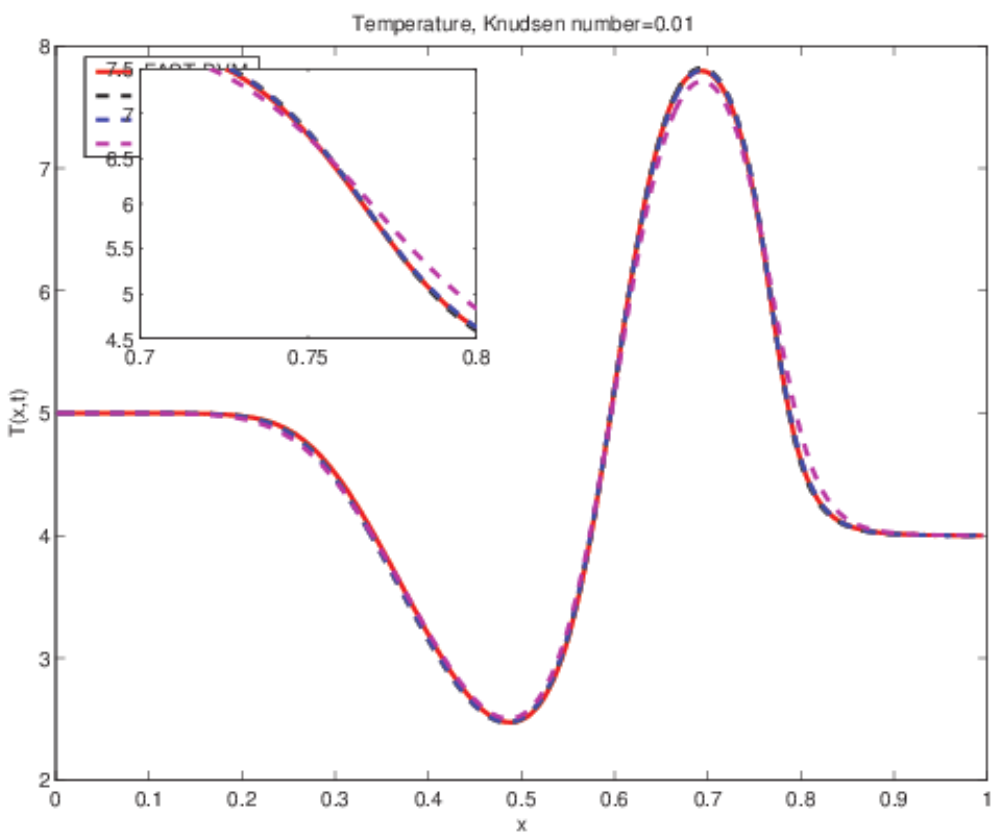}\\
\includegraphics[scale=0.43]{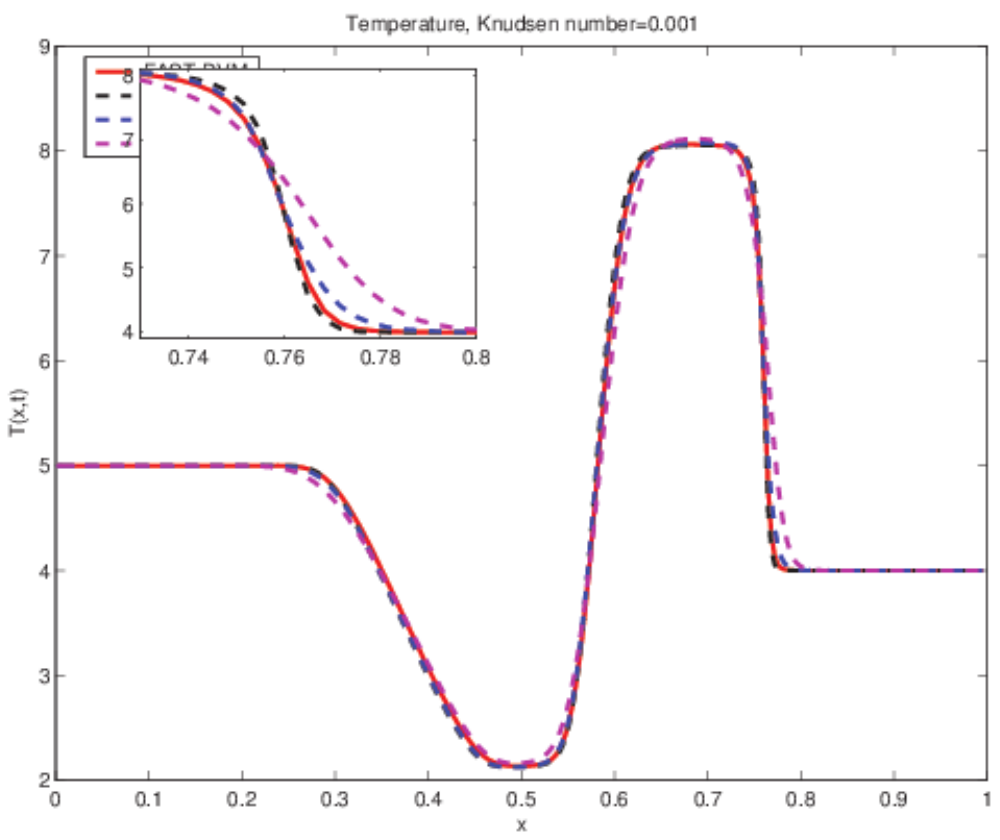}
\includegraphics[scale=0.43]{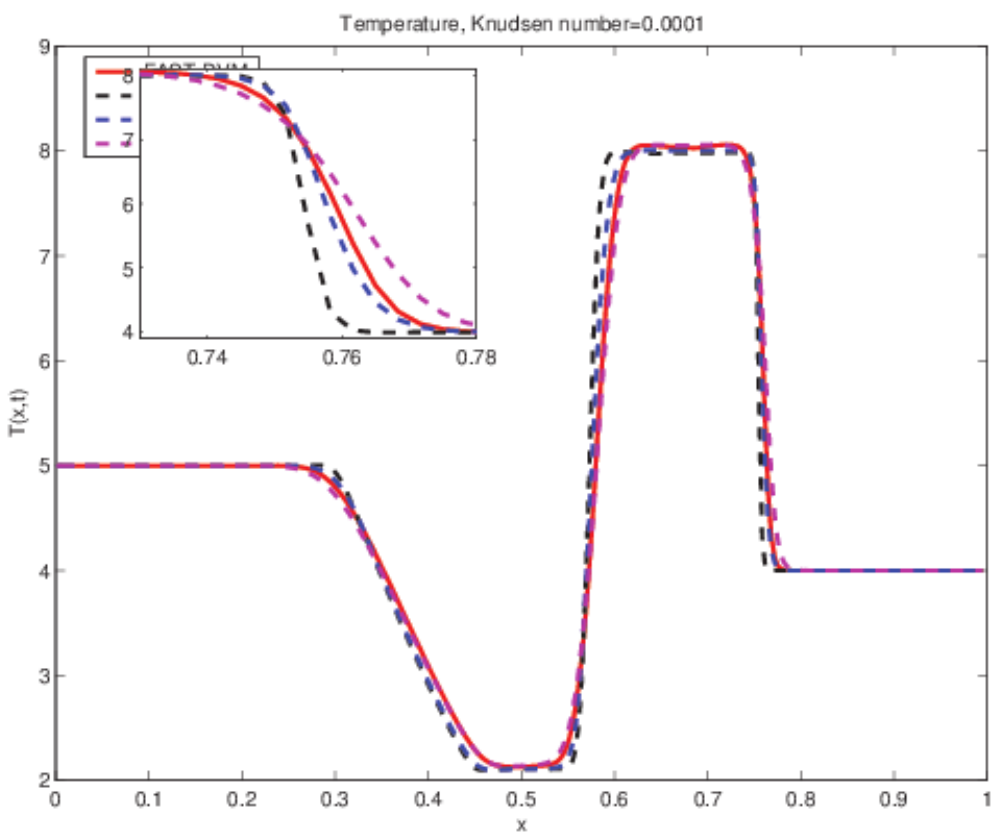}\\
\caption{1D Sod test: solution at $t_{\text{final}}=0.05$ for the
temperature, with $\tau=10^{-1}$ (top left), $\tau=10^{-2}$ (top
right), $\tau=10^{-3}$ (bottom left) and $\tau=10^{-4}$ (bottom
right).} \label{sod3}
\end{center}
\end{figure}

We consider the 1D/1D Sod test with $300$ mesh points in physical
and $100$ points in velocity spaces. The boundaries in velocity
space are set to $-15$ and $15$. The left and right states are given
by a density $\rho_{L}=1$, mean velocity $u_{L}=0$ and temperature
$T_{L}=5$ if $0 \leq x \leq 0.5$, while $\rho_{R}=0.125$, $u_{R}=0$,
$T_{R}=4$ if $0.5 \leq x \leq 1$. The gas is in thermodynamical
equilibrium. We repeat the same test with $4$ different values of
the Knudsen number, ranging from $\tau=10^{-1}$ to $\tau=10^{-4}$.
We plot the results for the final time $t_{\text{final}}=0.05$ for
the density (Figure \ref{sod1}), the mean velocity (Figure
\ref{sod2}) and the temperature (figure \ref{sod3}). In each figure
we compare the FKS method with a third order WENO method, a
second-order MUSCL method and a first-order upwind method
\cite{leveque:numerical-methods}. These numerical methods used as
reference, employ the same discretization parameters, except for the
time step which for stability reason is chosen equal to $\Delta t/2$
for the WENO and second-order MUSCL schemes, where $\Delta t$ is the
time step of the fast DVM method given by (\ref{eq:Time}).

\begin{figure}[h!]
\begin{center}
\includegraphics[scale=0.32]{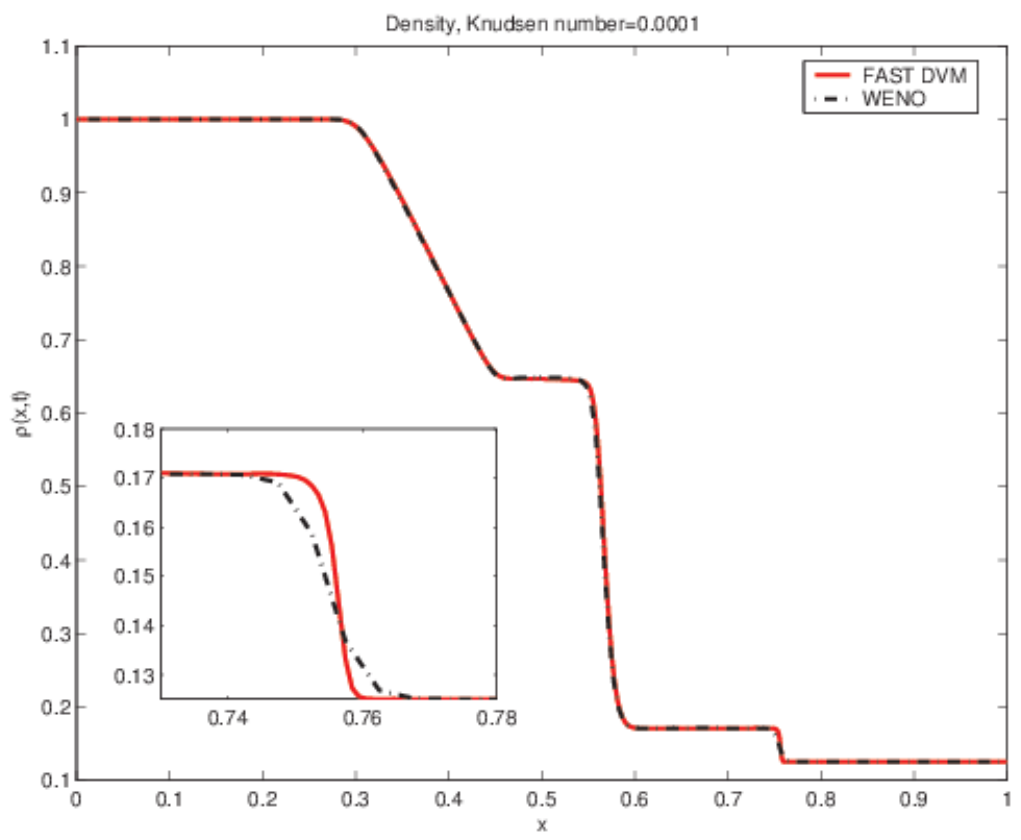}
\includegraphics[scale=0.32]{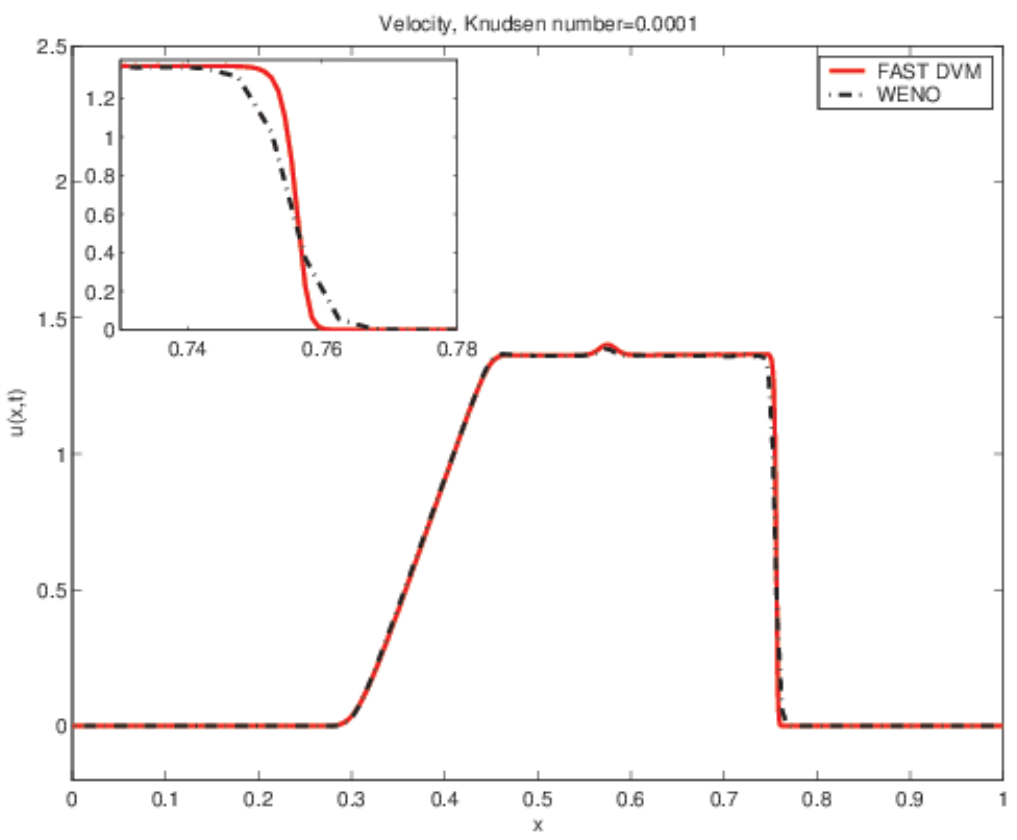}
\includegraphics[scale=0.32]{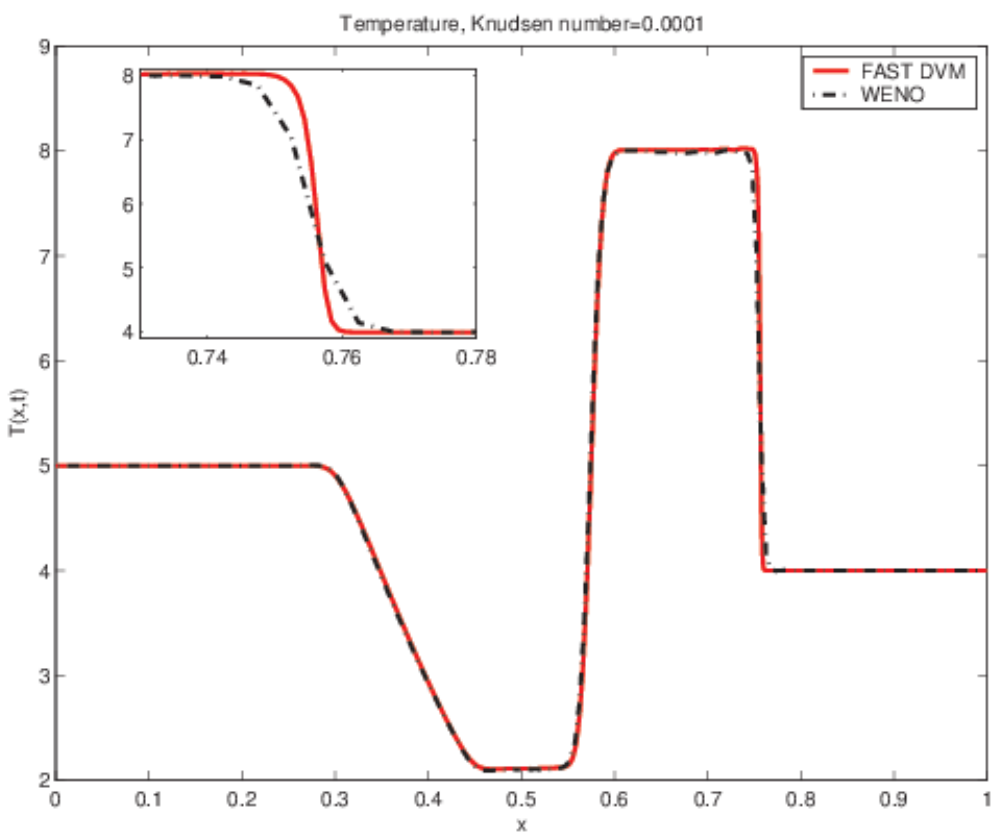}
\caption{1D Sod test: solution at $t_{\text{final}}=0.05$ for the
density, the mean velocity and the temperature with $\tau=10^{-4}$.
Comparison of solutions for the same computational time and
different meshes. WENO $200$ points (dashed line) and Fast DVM
$1000$ points (straight line).}\label{sod4}
\end{center}
\end{figure}

From Figures (\ref{sod1}) to (\ref{sod3}) we can observe that our
method gives very similar results to the two high order schemes for
$\tau=10^{-1}$, $\tau=10^{-2}$ and $\tau=10^{-3}$ while for
$\tau=10^{-4}$, the scheme is more diffusive than the second and
third order scheme but it still performs better than the first order
method. The behaviors of the method for different regimes are due to
the fact that for collisionless regimes the FKS gives almost the
exact solution, this means that it is more precise than the third
and second-order methods. When the gas becomes denser the projection
towards the equilibrium, which is only first-order (second step of
the method (\ref{relaxdvm})), does reduce the accuracy of the
method. Notice that high order reconstruction of the equilibrium
distribution could also be considered to increase the global
accuracy in such case. However, a key point of the FKS is its low
CPU time consumption in comparison to other existing methods. In the
case $\tau=10^{-4}$ for which the scheme exhibits diffusive
behaviors, a comparison between the third order WENO method and our
FKS method is carried out for a fixed CPU time. In other words, we
consider for a given total computational time, which method gives
better results. Thus, we solve the problem with $200$ points in
space and $100$ in velocity space for the WENO method and we
consider an FKS solver which employs $100$ points in velocity space.
In order to have the same computational time for the two methods, we
can afford $1000$ points for the FKS. The two results are compared
in Figure \ref{sod4}. We observe that, in this situation, the FKS
method gives more accurate solutions, in particular for the shock
wave (see the zooms in the figures). Finally, observe that the gain
in term of computational time is not so relevant for the one
dimensional case, while it becomes very important for the two and
the three dimensional case. In the later case, the difference is
about being able to do or not to do the computation in a reasonable
amount of time on a single processor machine.

\subsection{2D Sod shock tube problem}
\label{subsec_sod2D}

We consider now the 2D/2D Sod test on a square $[0,2]\times[0,1]$.
The velocity space is also a square with bounds $-15$ and $15$,
\textit{i.e.} $[-15,15]^{2}$, discretized with $N_v=20$ points in
each direction which gives $20^{2}$ points. We repeat the same test
using different $N_x\times N_y$ meshes ranging from $N_x=N_y=25$ to
$N_x=N_y=200$. The domain is divided into two parts,
a disk centered at point $(1,1)$ of radius $R_d=0.2$ is filled with a
 gas with density $\rho_{L}=1$, mean velocity $u_{L}=0$ and
temperature $T_{L}=5$, whereas the gas in the rest of the domain
is initiated with $\rho_{R}=0.125$, $u_{R}=0$, $T_{R}=4$.
The final time is $t_{\text{final}}=0.07$. The gas is in
thermodynamical equilibrium during all the computation which means
that we fix $\tau=0$. In practice, we are using the kinetic scheme
to compute the solution of the compressible Euler equation. We
recall that, as seen in the previous section, this is the case in
which the FKS scheme gives the worse results, this is due to the
first order accurate projection towards the local Maxwellian distribution.
However, this choice permits to compare our results with a numerical
method for the compressible Euler equations, being as already
stated, computationally very demanding to perform simulations of kinetic
equations in the two dimensional case and considerably more
demanding in the three dimensional case.

\begin{figure}
\begin{center}
\includegraphics[scale=0.43]{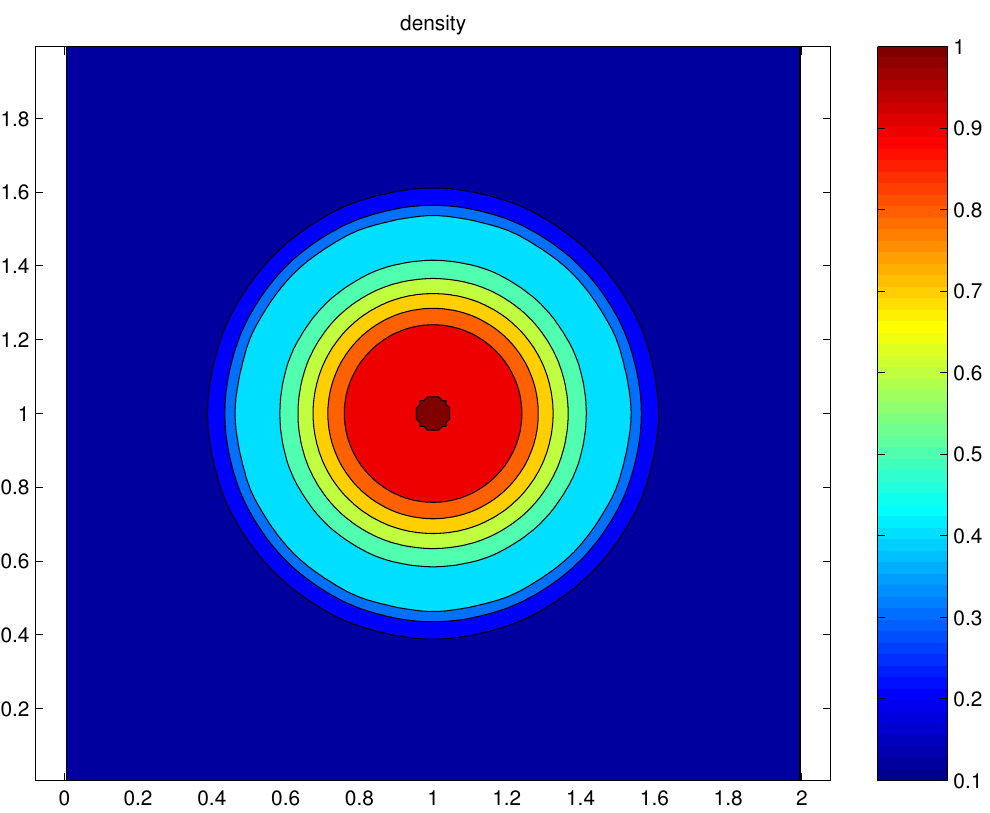}
\includegraphics[scale=0.43]{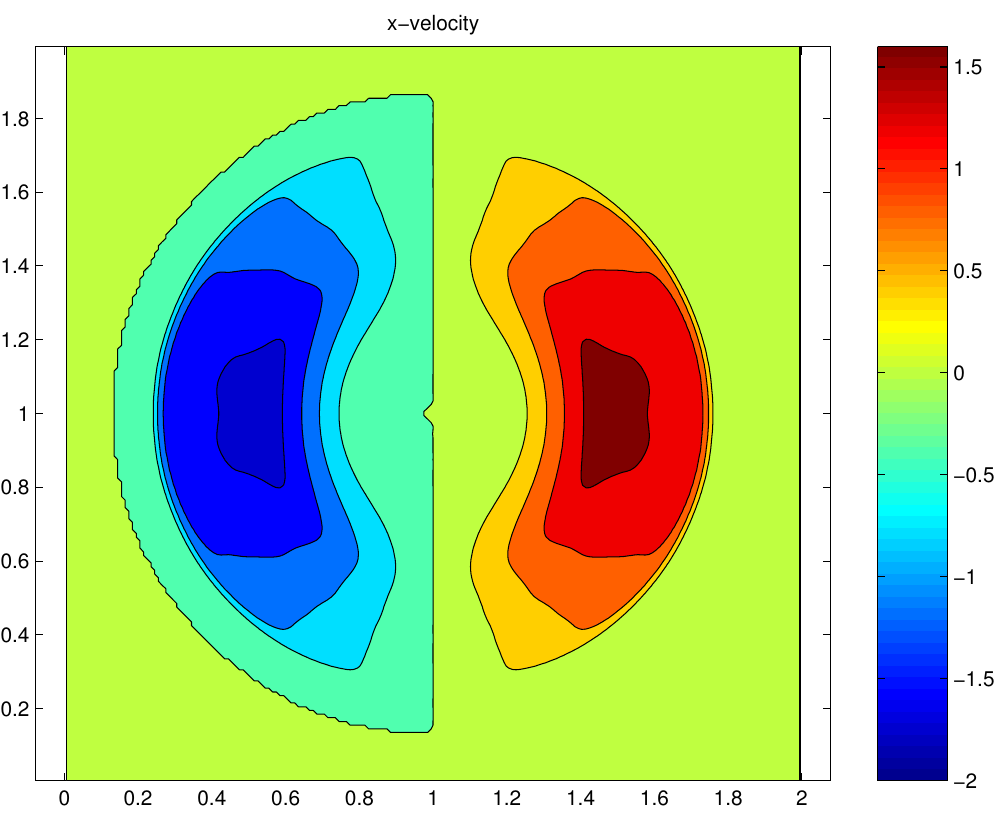}\\
\includegraphics[scale=0.43]{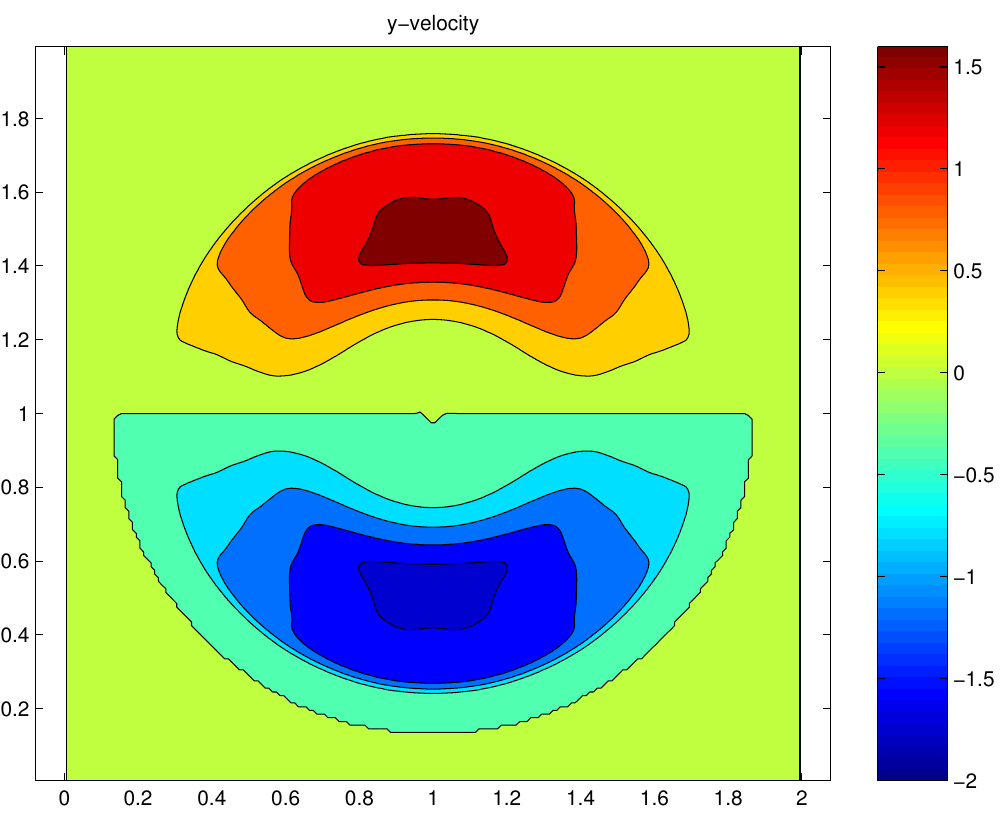}
\includegraphics[scale=0.43]{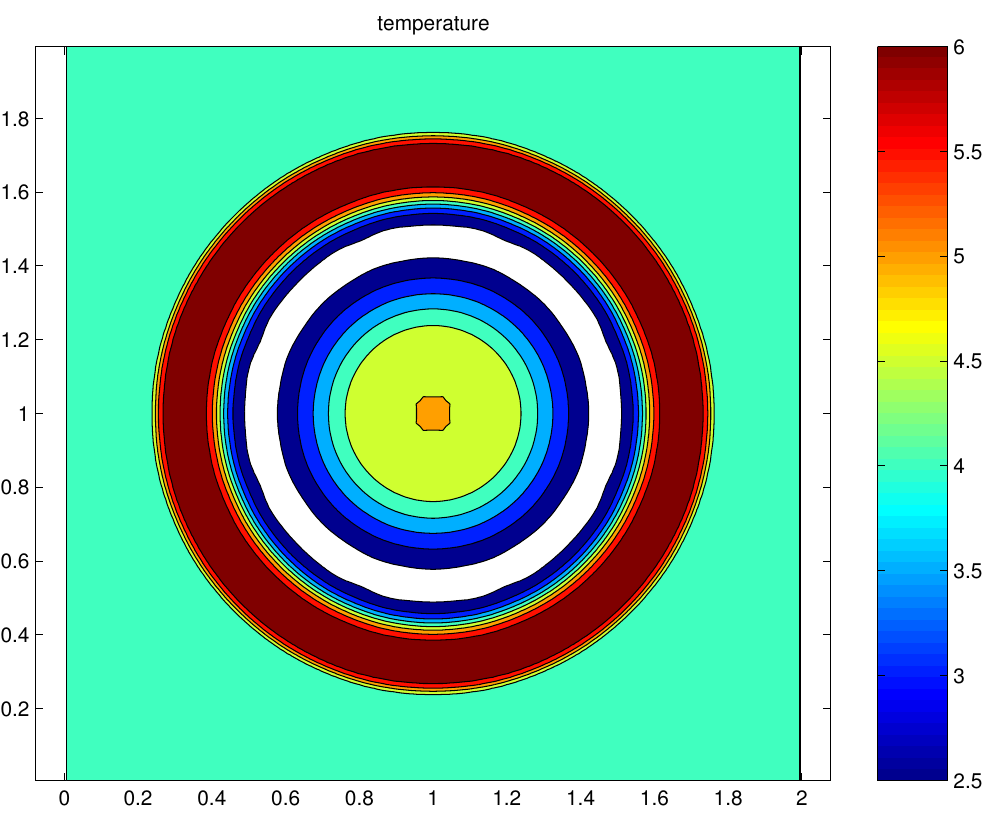}\\
\caption{2D Sod test: solution at $t_{\text{final}}=0.07$ for the
density (top left), the velocity in the x-direction (top right), the
velocity in the y-direction (bottom left) and the temperature
(bottom right).} \label{sod2D1}
\end{center}
\end{figure}

\begin{figure}
\begin{center}
\includegraphics[scale=0.4]{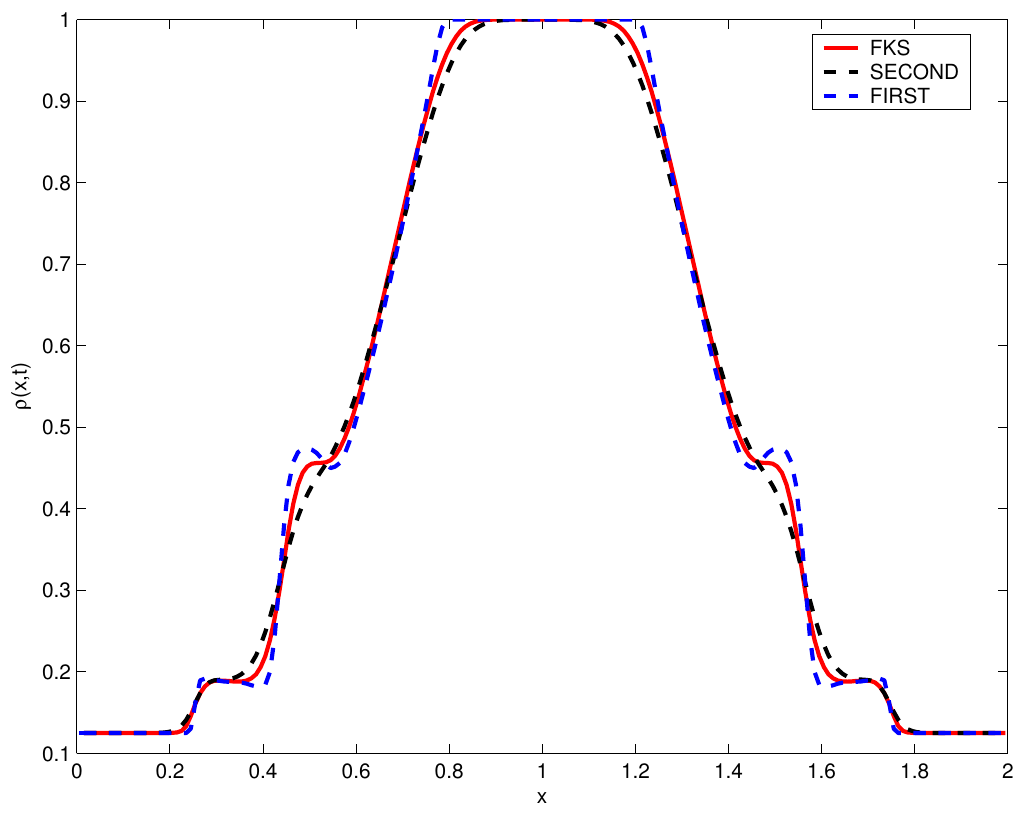}
\includegraphics[scale=0.4]{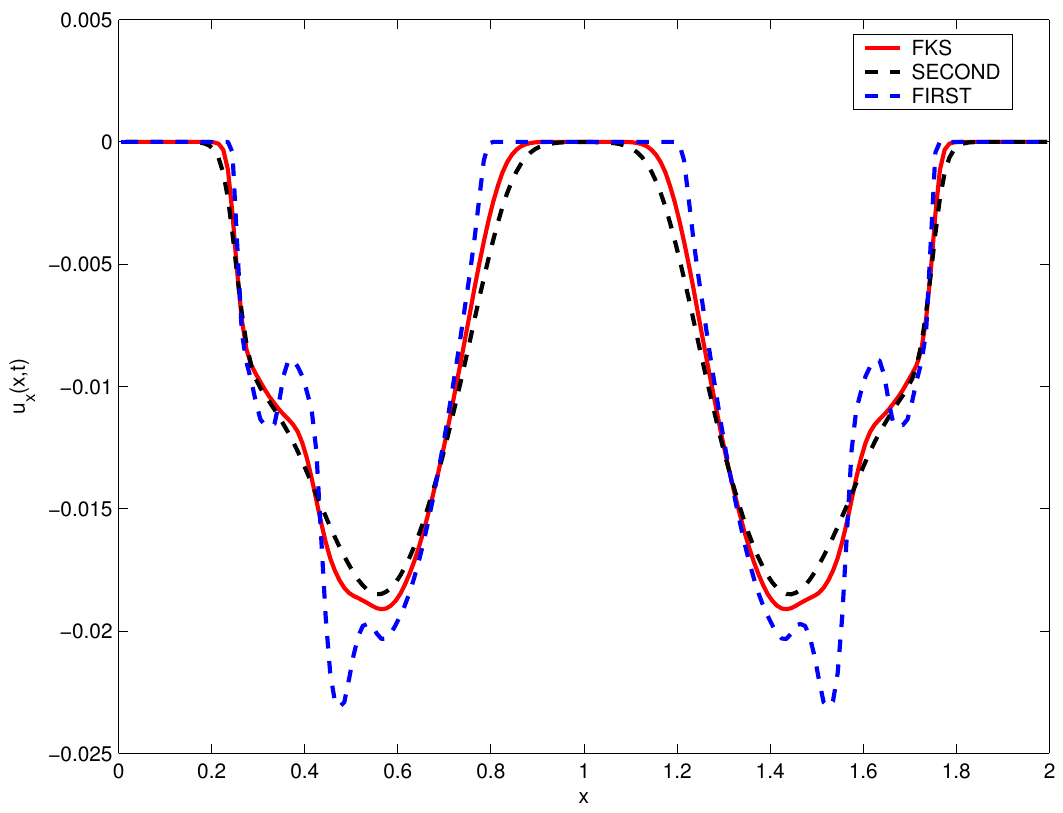}\\
\includegraphics[scale=0.4]{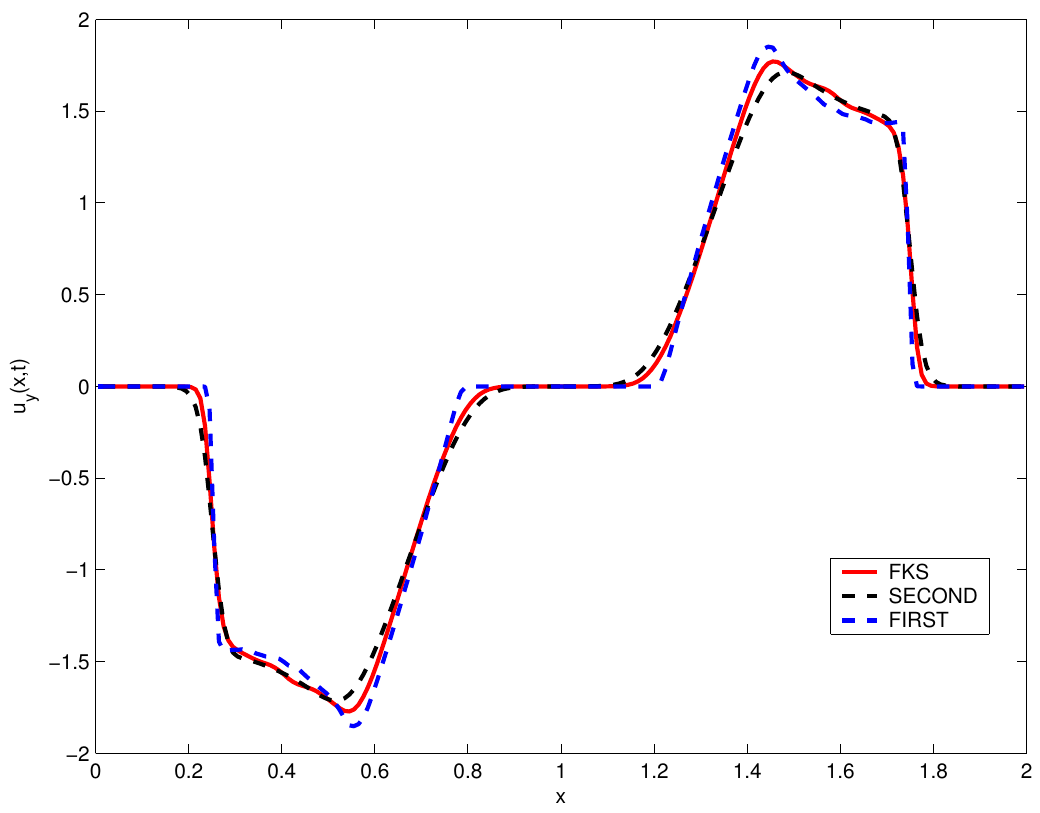}
\includegraphics[scale=0.4]{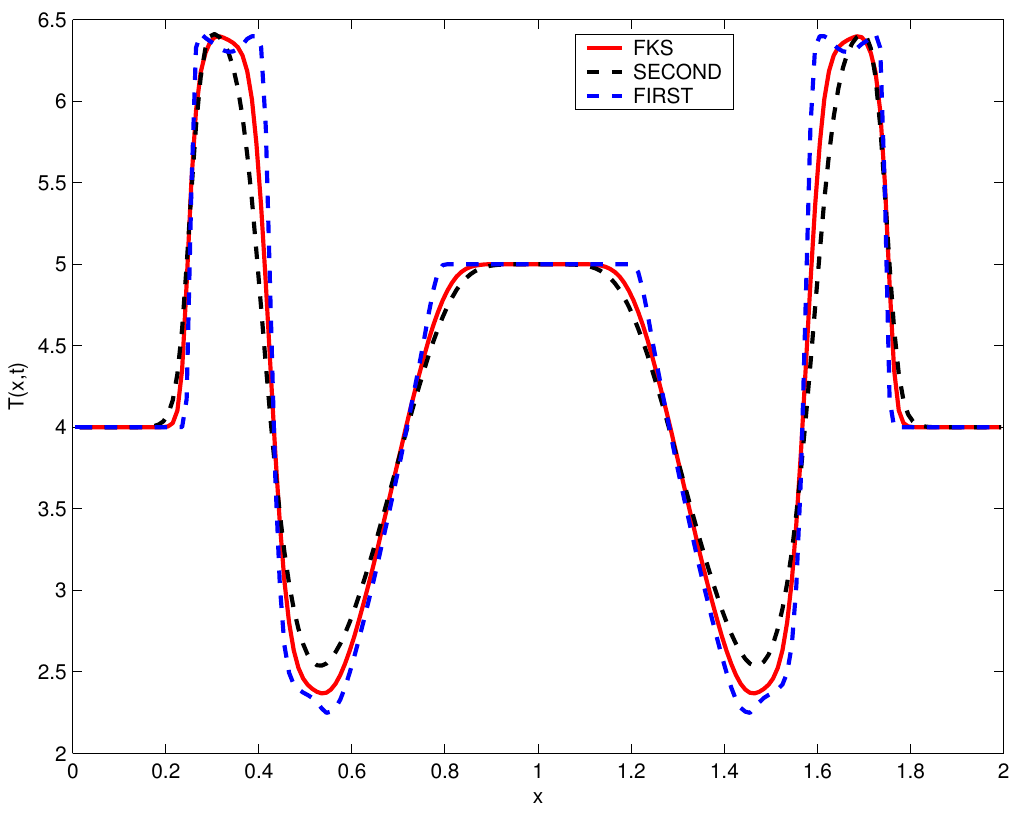}\\
\caption{2D Sod test: solution (continuous line) at
$t_{\text{final}}=0.07$ and $x=1$ for the density (top left), the
velocity in the x-direction (top right), the velocity in the
y-direction (bottom left) and the temperature (bottom right).
Comparisons with first order and second order MUSCL methods (dotted
lines)} \label{sod2D2}
\end{center}
\end{figure}

In Figure~\ref{sod2D1} we show the results for respectively the
density, the mean velocity in the $x$-direction and in the $y$-direction
and the temperature using a $200\times 200$ mesh. In
Figure~\ref{sod2D2} we report the profile for $x=1$ of the same macroscopic
quantities comparing the results to a first order and to a second
order MUSCL scheme for the compressible Euler equations
\cite{leveque:numerical-methods}. We clearly see that, as in the 1D
case, the accuracy of the FKS method lies between the first and the
second order accuracy in the limit $\tau\rightarrow 0$. We expect
the accuracy to be highly improved when the gas is far from the
thermodynamical equilibrium as in the one dimensional case.

In table~\ref{tab:CPU_Sod2D} we report the CPU time $T$ of these
simulations, the CPU time per time cycle $T_{\text{cycle}}$, the CPU
time per cycle per cell $T_{\text{cell}}$ and the number of cycles
needed to perform the computation for different meshes in space and
a fixed mesh in velocity. As expected the number of time step
linearly scales with the size of the spatial mesh at fixed velocity
mesh (factor $2$ when the cell number is multiplies by $4$). The CPU
time is very small compared to classical kinetic schemes, in less
than $10$ minutes the simulation of the Sod shock tube on a
$200\time 200$ mesh is computed. Finally we observe that the CPU
time per cycle per cell is almost constant which allows to predict
the end of the simulation and its cost beforehand.

%
%

\subsection{Numerical validation of the 3D/3D fast FKS method}

Here we report some simulations of the full 3D/3D problem. We
consider only the case in which $\tau \equiv 0$, which means, we
project towards equilibrium at each time step, this is the fluid
limit. We recall that, in this regime, the numerical method gives
the worst results in terms of precision, on the other hand, exact
solution are known and this permits to make fair comparisons. For
all the other regimes, the performances of the method are better as
shown in the previous section.

The FKS method has been implemented in fortran on a sequential
machine. The goal is to numerically show that such a kinetic scheme
can reasonably perform on six dimensions on a mono-processor laptop.
All simulations have been carried out on a HP EliteBook 8740W
Intel(R) Core(TM) i7 Q840@1.87GHz running under a Ubuntu (oneiric)
version 11.10. The code has been compiled with gfortran 4.6 compiler
with -O3 optimization flags.

Otherwise noticed the velocity space is $[-15,15]^3$ or $[-10,10]^3$
and is discretized with $N_v=13$ or $N_v=12$ grid points in each
velocity direction, leading to $N_v^3= 2197$ or $1728$ mesh points.
The time step is fixed to $95\%$ of the maximum time step allowed,
as prescribed by the CFL condition (\ref{eq:Time}), apart from the
last time step which is chosen to exactly match the user-given final
time. Symmetric boundary conditions are considered.

The Sod shock tube in 1D is run as a sanity checks in order to
validate the implementation of the method and show its ability to
reproduce 1D results with a 3D run. Then the Sod problem
in 3D is simulated to show the performances of the
FKS algorithm and further compared to a reference solution.
For each simulation we report the memory
consumption, the full CPU time and the CPU time cost per cell per
time step. Some extrapolation of these results are also made to
measure the efficiency of this method.

\subsubsection{1D Sod shock tube problem: A sanity check}

\begin{table}
  \begin{tabular}{|cc|c|c||c|c|c|c|}
    \hline
     \multicolumn{2}{|c|}{\textbf{Cell v \#}}  &\textbf{Cell x \#}  $N_c$ & \textbf{Cell $x\times v$ \#}  $N_{tot}$ & \textbf{Cycle} & \textbf{Time} & \textbf{Time/cycle} & \textbf{Time/cell} \\
    $N^{2}_v$ & Bounds &$N_x\times N_y$ & $N_x\times N_y\times K^{2}$ &  $N_{\text{cycle}}$  &  $\text{T}$ (s)  & $T_{\text{cycle}}$ (s) & $T_{\text{cell}}$ (s) \\
    \hline
    \hline
    &&
    $25 \times 25$ & $25 \times 25\times 20^2$ & $13$   & $2$s   & $0.1538$  & $2.46\times 10^{-4}$ \\
    &&$=625$ & $=250000$ & & & & \\
    \cline{3-8}
    \multirow{4}{*}{ $20^2$} &
    \multirow{4}{*}{\begin{sideways} $[-15,15]$  \end{sideways}}
    &$50 \times 50$ & $50 \times 50\times 20^{2}$ & $25$   & $8$s   & $0.32$  & $1.28\times 10^{-4}$ \\
    &&$=2500$ &$=10^6$ & &  & & \\
    \cline{3-8}
    &&$100 \times 100$ & $100 \times 100 \times 20^{2}$  & $50$  & $60$s   & $1.2$ & $1.20\times 10^{-4}$ \\
    &&$=10000$ &$=4 \ 10^6$ & & $1$mn &  & \\
 \cline{3-8}
    &&$200 \times 200$ & $200 \times 200\times 20^{2}$ & $100$  & $490$s   & $535.75$ & $1.22\times 10^{-4}$ \\
    &&$=40000$ &$=16 \ 10^6$ & & $\sim 8$mn &  & \\
    \hline
\end{tabular}
\caption{ \label{tab:CPU_Sod2D}
    2D Sod shock tube.
    The time per cycle is obtained by $T_{\text{cycle}} = \text{T}/N_{\text{cycle}}$ and
    the time per cycle per cell by $T_{\text{cell}} = \text{T}/N_{\text{cycle}}/N_c$.
}
\end{table}

The first sanity check consists of running the 1D Sod shock tube in
$x$ direction on $N_x \times 2 \times 2$ cubes. The initial data are
the same as for the 1D problem previously run. The final time is
$t_{\text{final}}=0.1$. In our numerical experiments the
computational domain is of size $1$ in $x$ direction leading to
$\Delta x = 1/N_x$. We set $\Delta y = \Delta z = \Delta x$. Four
successively refined meshes in $x$ direction are utilized, $N_x=50,
100$, $200$, and $400$, in order to observe the convergence of the
numerical method towards the exact solution.

\begin{figure}
  \begin{center}
    \begin{tabular}{cc}
      \includegraphics[width=0.45\textwidth]{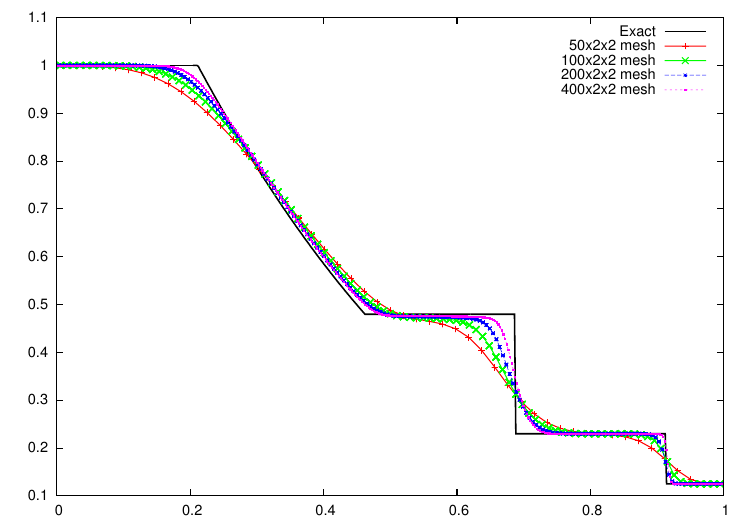}
      &
      \includegraphics[width=0.55\textwidth]{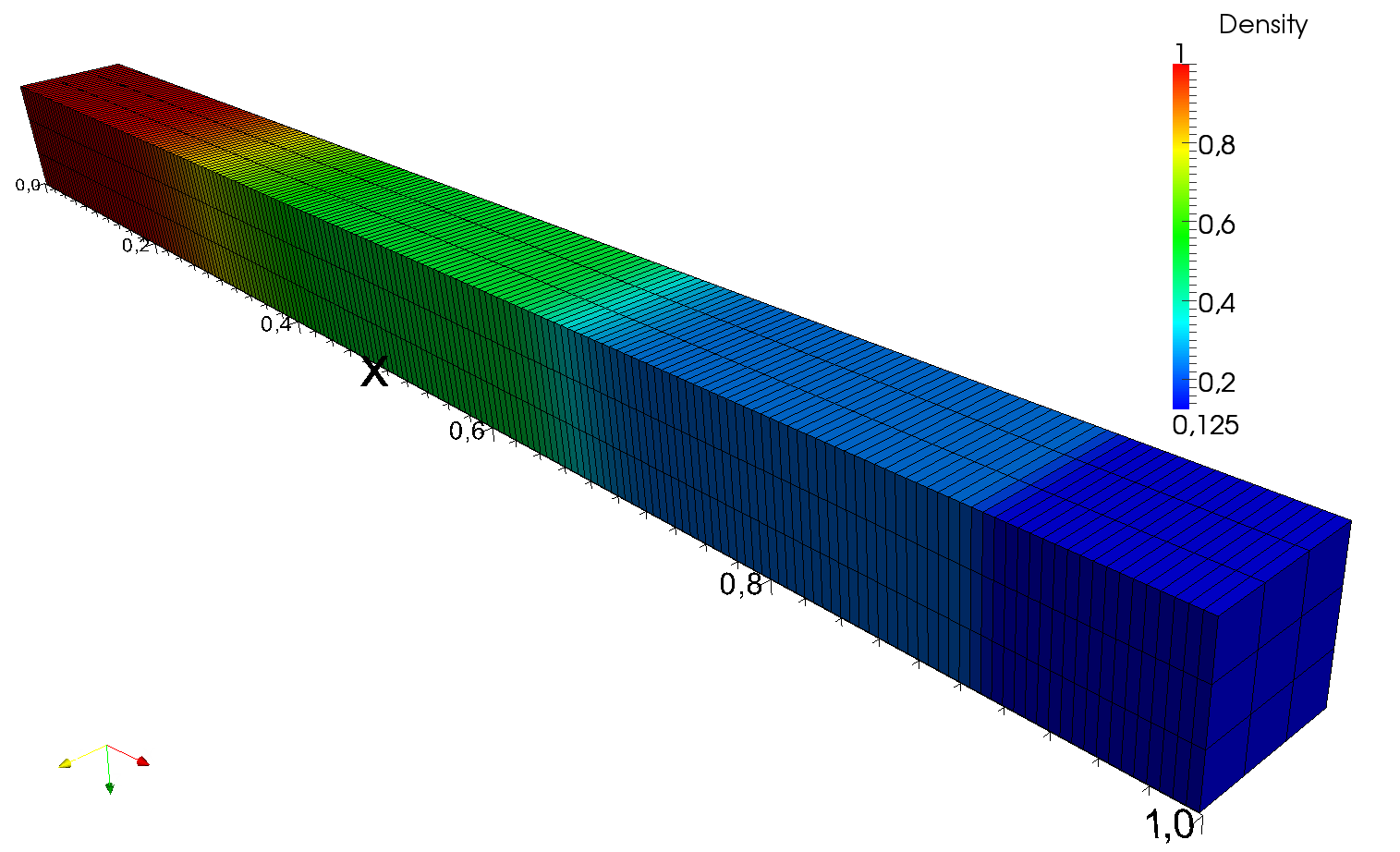}
      \\
      \textbf{(a)} & \textbf{(b)} \\
      \includegraphics[width=0.45\textwidth]{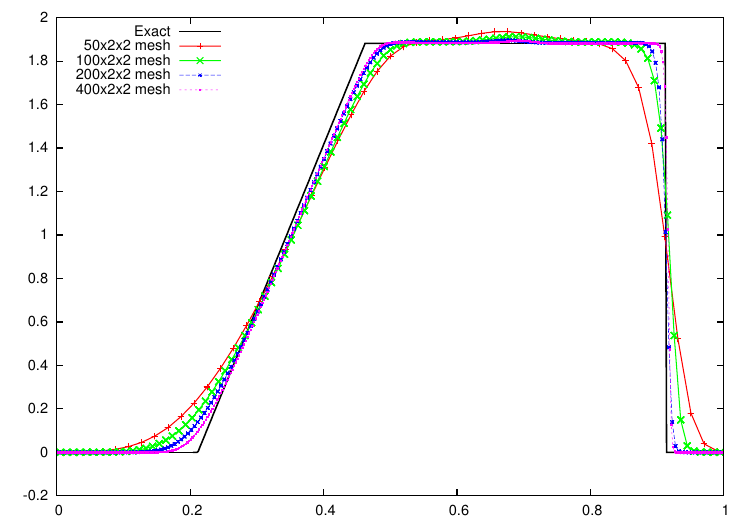}
      &
      \includegraphics[width=0.45\textwidth]{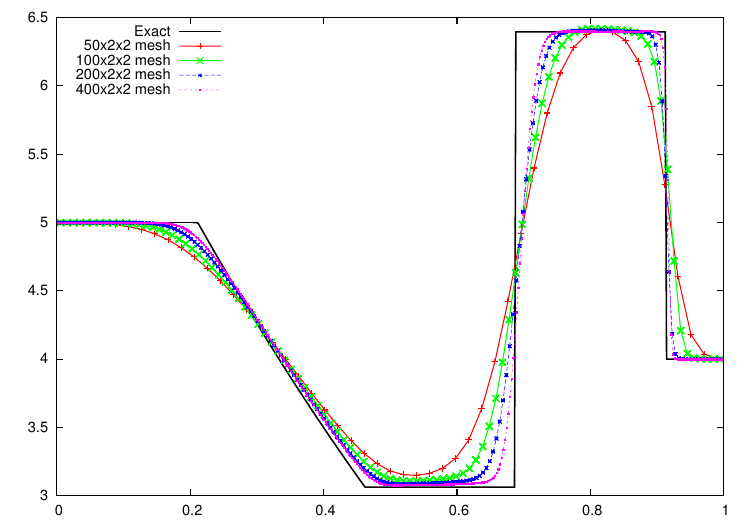}
      \\
      \textbf{(c)} & \textbf{(d)} \\
    \end{tabular}
    \caption{ \label{fig:sod1D}
      3D-1D Sod problem at $t_{\text{final}}=0.1$ for $50, 100$, $200$, and $400$ cells
      in $x$ direction and $2$ in $y$ and $z$ directions ---
      Panels \textbf{(a)}, \textbf{(c)}, \textbf{(d)}: Density, velocity and temperature as a function of $x$
      \textit{vs} exact solution (straight line) ---
      Panel \textbf{(b)}: 3D view of colored density for a $200 \times 3 \times 3$ mesh. }
  \end{center}
\end{figure}

In Figure~\ref{fig:sod1D} we display the density, the velocity and the
temperature {\it vs} the exact solution with solid line
(respectively panels \textbf{(a)}, \textbf{(c)} and \textbf{(d)})
and a 3D view on the mesh cells colored by density (panel
\textbf{(b)} where a $200 \times 3 \times 3$ mesh is used for figure
scaling reasons). The first observation is the perfect symmetry in
the ignorable directions $y$ and $z$ as all cells are plotted
(notice that the results for a $N_x \times 5\times 5$ cells mesh
exactly match the $N_x \times 2\times 2$ results). The second
obvious observation is the convergence of the numerical solution
towards the exact solution when the mesh is refined. These results
assess the ability of the method and the code to reproduce 1D
results without alteration.

In table \ref{tab:CPU_Sod1D} we gather the number of cycles
$N_{\text{cycle}}$, the CPU time $T$ of these simulations and
display the CPU time per time cycle $T_{\text{cycle}}$ and the CPU
time per cycle per cell $T_{\text{cell}}$. As expected, the cycle
number and the CPU time per time cycle scales with the cell number
and, consequently, the CPU time per cycle per cell is almost
constant. This allows to almost exactly predict the duration of a
simulation knowing the cell number. Moreover we have provided the
relative percentage of the cost of the transport and collision
stages.

As expected the transport stage does not cost anything, in absolute
value, especially when the number of cells increases. In fact, for
computing the solution of this stage in all domain, we consider the
evolution of the distribution function $f$ in one single cell, the
same happens in the other cells. This means that the cost of this
stage is proportional to the $N_v^{3}$ mesh points in the velocity
space. On the other hand, in finite volume methods as well as Monte
Carlo method the cost to solve this stage is proportional to
$N_v^{3}N_c$ with $N_c=N_{x}N_{y}N_{z}$ and, obviously this scales
with $N_c$.
Another satisfactory result is the memory storage $Mem$ in MB (or
Mo) of the method which is very low because we never have to store
the distribution function values for more than $N_v^{3}$ points,
leading to store $13^3 \times 7$ reals, say $\sim 0.123$MB
independently of $N_c$.
 Conversely the storage of the Monte Carlo method
scales with the cell number $N_c$. Finally as expected the time
$\text{T}$ scales with a factor $4$ for twice the number of cells.
\begin{table}
  \begin{tabular}{|c||c|c|cc|c|c|}
    \hline
    \textbf{Cell \#}   & \textbf{Cycle} & \textbf{Time} & \multicolumn{2}{c|}{\textbf{Time/cycle}} & \textbf{Time/cycle/cell} & \textbf{Memory}\\
    $N_x\times N_y\times N_z\times N^{3}_v$ &  $N_{\text{cycle}}$  &  $T$(s)  & \multicolumn{2}{c|}{$T_{\text{cycle}}$(s)}   & $T_{\text{cell}}$(s) & Mem(MB)  \\
    $=N_c \times N^{3}_v$ & & & \textit{Transp.} & \textit{Coll.} & & \\
    \hline
    \hline
    $50 \times 2 \times 2$  & $81$   & $18$   & \multicolumn{2}{c|}{$0.22$} & $1.11\times 10^{-3}$ & $0.660$\\
    $=200\times 13^3=439400$ & & & $0.07 \%$ & $99.93 \%$ & &\\
    \hline
    $100 \times 2 \times 2$ & $160$  & $68$   & \multicolumn{2}{c|}{$0.43$} & $1.07\times 10^{-3}$ & $0.704$\\
    $=400\times 13^3=878800$ & & & $0.05 \%$ & $99.95 \%$ & & \\
    \hline
    $200 \times 2 \times 2$ & $318$  & $276$  & \multicolumn{2}{c|}{$0.87$} & $1.08\times 10^{-3}$  & $0.812$\\
    $=800\times 13^3=1757600$ & & & $0.03 \%$ & $99.97 \%$ & &\\
    \hline
    $400 \times 2 \times 2$ & $634$  & $1071$  & \multicolumn{2}{c|}{$1.69$} & $1.06\times 10^{-3}$ & $1.000$\\
    $=1600\times 13^3=3515200$ & & & $0.02 \%$ & $99.98 \%$ & &\\
    \hline
  \end{tabular}
  \caption{ \label{tab:CPU_Sod1D}
    1D Sod shock tube run with the 3D/3D FKS method.
    The time per cycle is obtained by $T_{\text{cycle}} = \text{T}/N_{\text{cycle}}$ and
    the time per cycle per cell by $T_{\text{cell}} = \text{T}/N_{\text{cycle}}/N_c$.
    The relative percentage of the cost of the transport and relaxation
    stages are provided.
    For our FKS method the transport stage costs almost nothing.
  }
\end{table}

 \subsubsection{3D Sod shock tube problem}
\begin{figure}
   \begin{center}
     \includegraphics[width=0.7\textwidth]{figure/SOD/sod3D_rho1D_comp_mix}
     \begin{tabular}{cc}
       \includegraphics[width=0.45\textwidth]{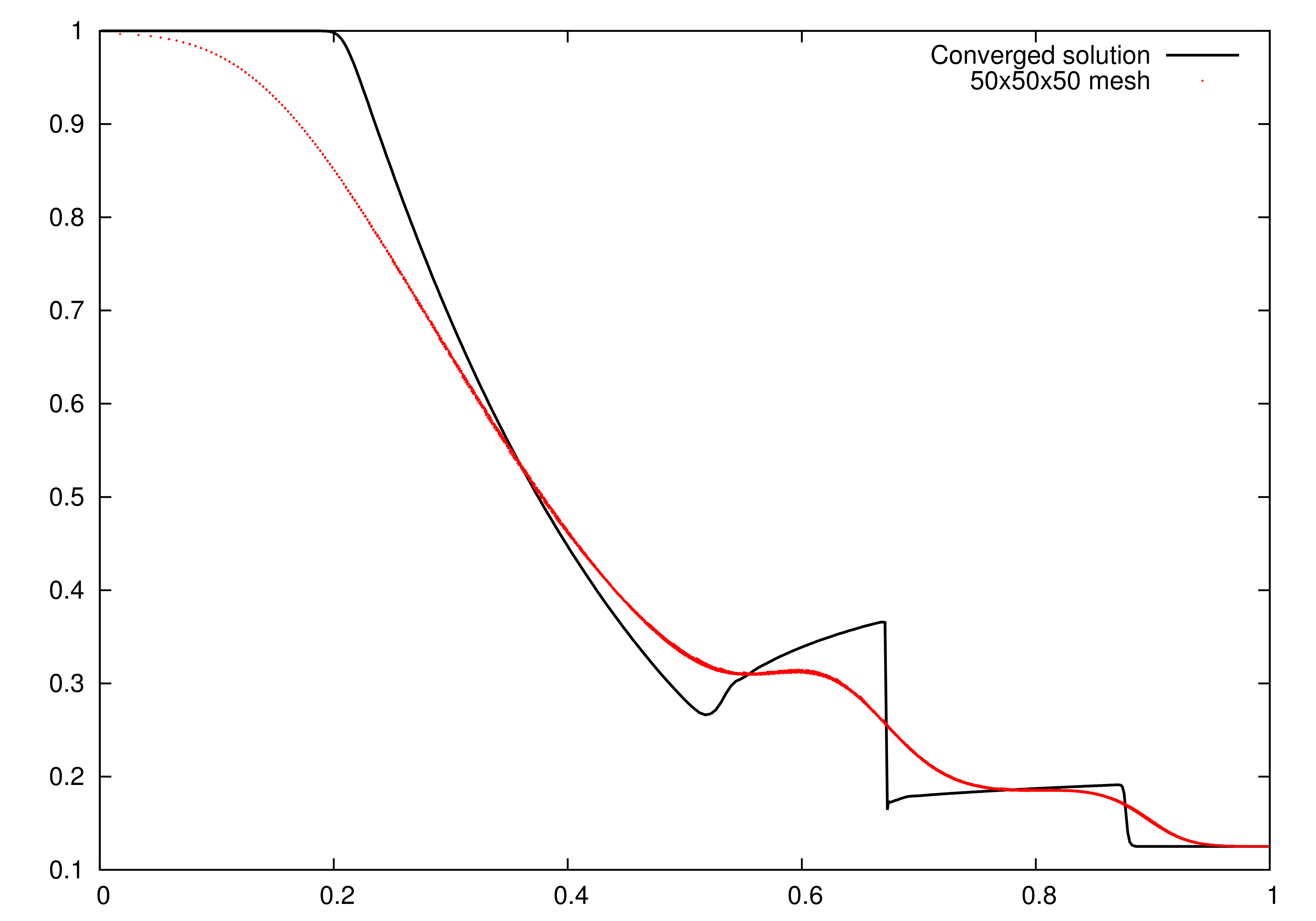}
       &
       \includegraphics[width=0.45\textwidth]{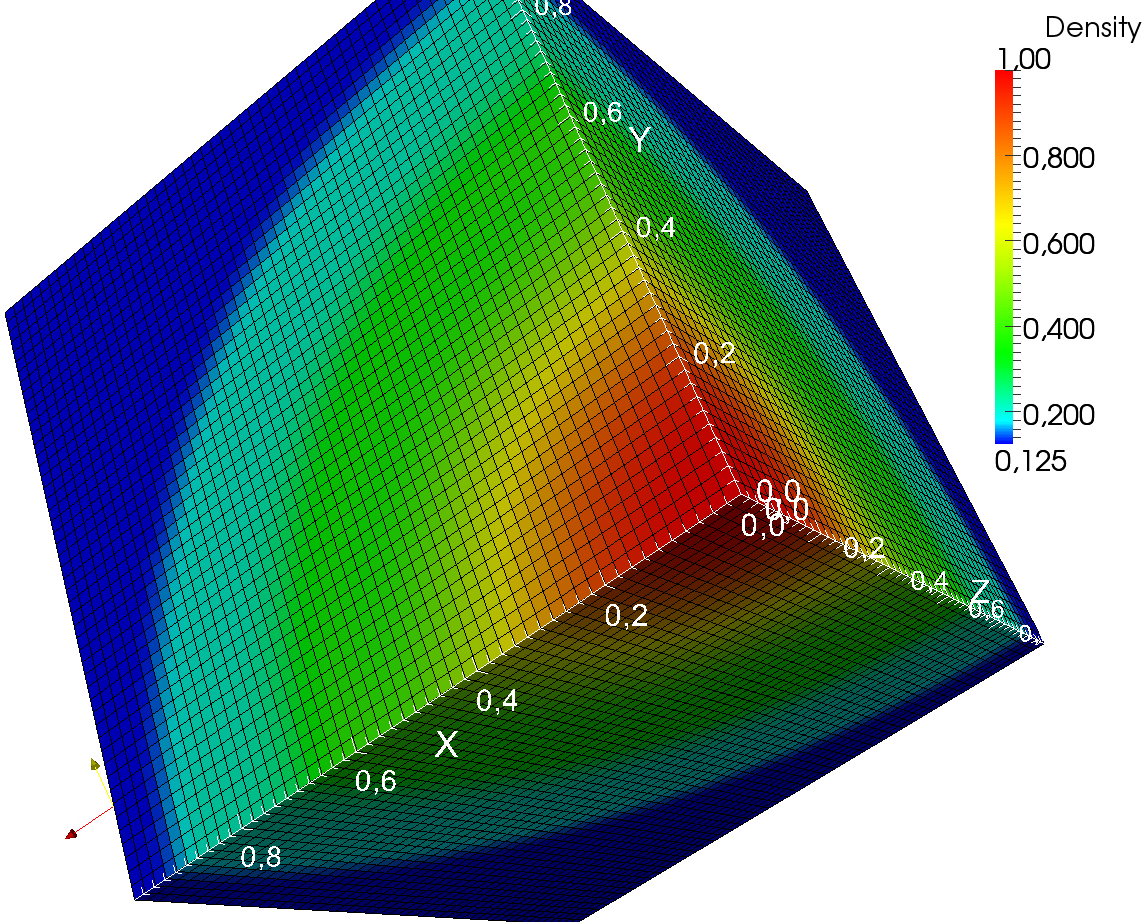}
       \\
       \includegraphics[width=0.45\textwidth]{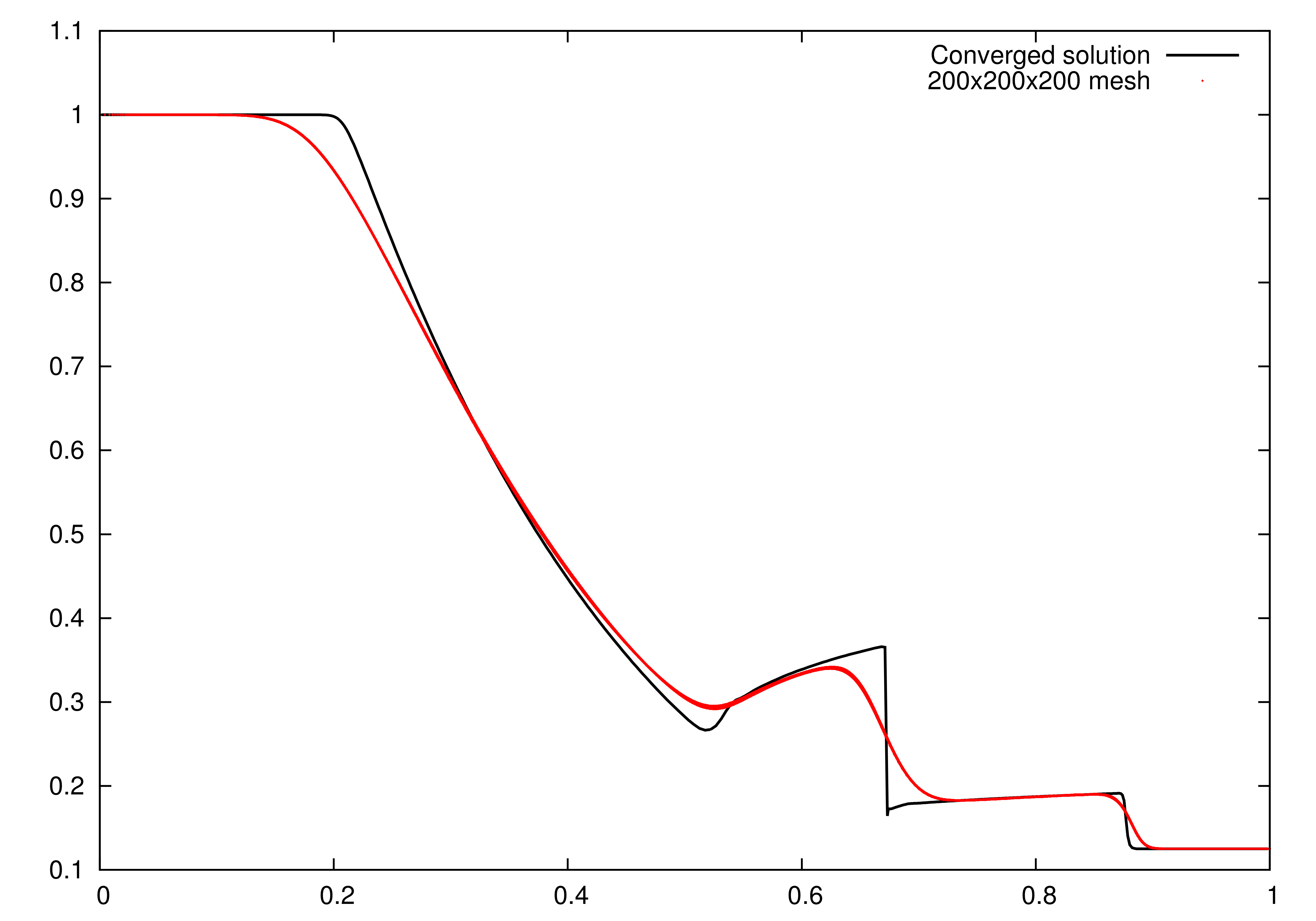}
       &
       \includegraphics[width=0.45\textwidth]{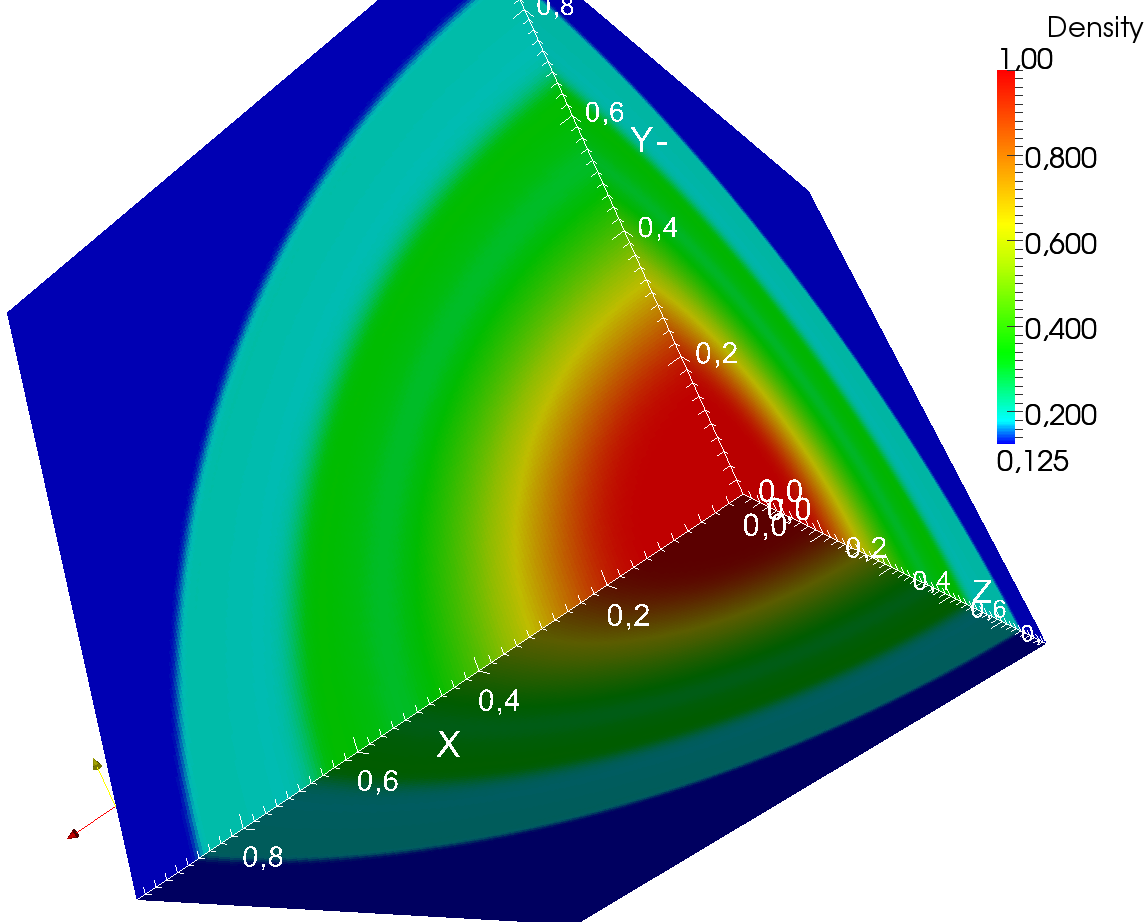}
       \\
     \end{tabular}
     \caption{ \label{fig:sod3D_2}
       Sod problem at $t_{\text{final}}=0.1$ for $N_x\times N_x\times N_x$ cells (for $N_x=50,100, 200$)
       for the velocity space $[-10;10]$ discretized with $12^{3}$ mesh points.
       ---
       Top:  Convergence of density as a function of cell center radius for all cells
       \textit{vs} converged solution
       (straight thick line) for the three meshes with zooms on contact and shock waves.
       Left: Density as a function of cell center radius (middle: $N_x=50$, bottom: $N_x=200$)
       Right: 3D view of density on the unit cube $N_x=50$ (middle) and $N_x=200$ (bottom)
       (the mesh is only shown for $N_x=50$). }
   \end{center}
 \end{figure}

 The 3D Sod shock tube has been run with the 3D/3D FKS method.
 The left state of the 1D Sod problem
 is set for any cell $c$ with cell center radius $r_c\leq 1/2$, conversely the
 right state is set for cell radius $r_c > 1/2$.  The final time is
 $t_{\text{final}}=0.1$.
 The domain is the unit cube and
 the mesh is composed of $N_x\times N_x \times N_x$ cells with $\Delta x = 1/N_x$ and
 $\Delta x = \Delta y=\Delta z$.
 The problem is run with $N_x=50$ ($125 000$ cells), $N_x=100$ ($1$ million cells)
and $N_x=200$ ($8$ millions cells).
 The velocity space is either $[-10;10]$ discretized with $12^{3}$ points,
 or $[-15;15]$ discretized with $13^{3}$ points.
This leads to consider up to $200^{3}\times 13^3\simeq 17.7$
milliards cells.
In Figure~\ref{fig:sod3D_2} the density is plotted as a function of
the radius (left panel) and the colored density on a 3D view (right
panel) for $N_x=50$ (middle panels) and $N_x=200$
(bottom panels). The
two different choices for the bounds and the mesh points in velocity
space do not significantly change the results hence only the
solution with bounds $[-10;10]$ and with $12^{3}$ mesh points is
reported.
The reference solution is obtained with a 2D axisymmetric compatible staggered
Arbitrary-Lagrangian-Eulerian code \cite{ALE_INC} with $1000$ cells in radial
and $20$ cells in angular directions.
Moreover in Figure~\ref{fig:sod3D_2} (top panel) we present the convergence of the density
as a function of cell center radius for all cells for the
$50\times 50\times 50$, $100\times 100\times 100$  and $200\times 200\times 200$
cells meshes. These curves are compared to the reference solution
in straight thick line and they show that the results are converging towards the reference solution.
In table~\ref{tab:CPU_Sod3D_bis} we gather the number of time steps
and the total CPU time $\text{T}$ for $50^{3}$ and $100^{3}$ cell
meshes for the two different configurations: one with $N_v=13$ and
the velocity space $[-15,15]$ and the second one with $N_v=12$ and
the velocity space $[-10,10]$. For the $50^{3}$ mesh the simulation
takes $45$ minutes or $1.36$ hour depending on the configuration.
For the finer $100^{3}$ mesh the simulation takes either $11$ hours
or $24$ hours The memory consumption ranges from $124$Mb to $924$Mb
depending on the configurations and it scales with the number of
cells $N_c$.

\begin{figure}
  \begin{center}
    \begin{tabular}{cc}
      \includegraphics[width=0.485\textwidth]{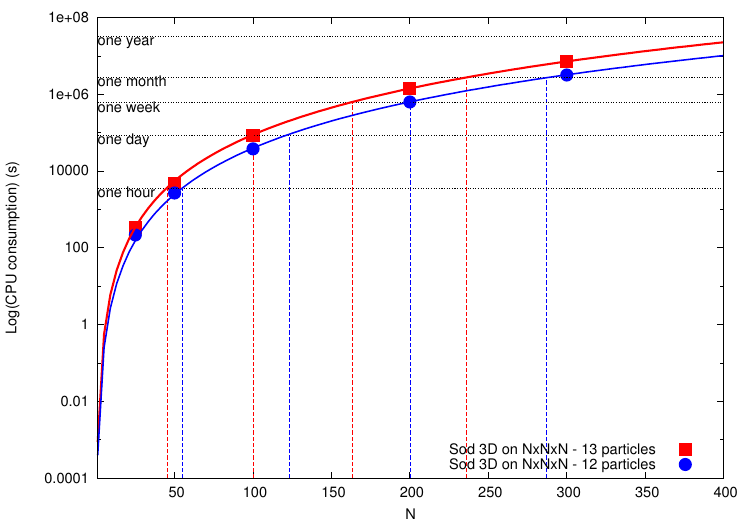}
      &
      \includegraphics[width=0.485\textwidth]{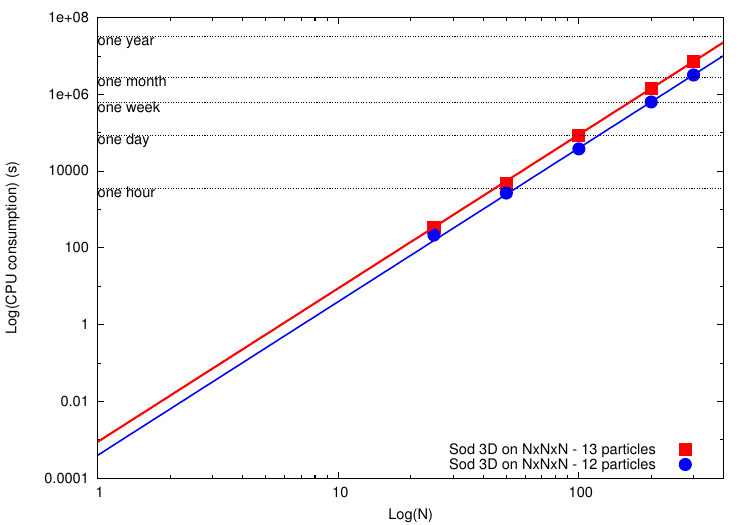}
    \end{tabular}
    \caption{ \label{fig:sod3D_CPU}
      Left: Log of the CPU time consumption for the 3D Sod problem at $t_{\text{final}}=0.1$
      as a function of $N$ (for $N\times N\times N$ cell meshes) on a single processor laptop
      The red/blue squares are taken from Table~\ref{tab:CPU_Sod3D_bis}, the thick red/blue curves are
      the extrapolation curve from $T_{\text{cell}}$. The horizontal lines corresponding to one hour, one
      day, week, month and year are also plotted. $N=100$ corresponds to the 'one million cells' in space  ---
      Right: Log/Log scale.
    }
  \end{center}
\end{figure}

Then, we compute the cost per cycle $T_{\text{cycle}}$ and per cycle
per cell $T_{\text{cell}}$. One observe that the cost per cycle per
cell is an almost constant equal to $4\times 10^{-4}$s or $5.5\times
10^{-4}$s. The extrapolation of the CPU time $T$ for a $200^{3}$
mesh at $T_{\text{cell}}$ fixed leads to one or two weeks
computation for the two configurations and a memory storage of about
$900$MB.
\begin{table}
  \begin{tabular}{|cc|c||c|c|c|c|c|}
    \hline
        &  &\textbf{Cell \#} $N_c\times N_v^{3}$    & \textbf{Cycle} & \textbf{Time} & \textbf{Time/cycle} & \textbf{Time/cell} & \textbf{Mem}\\
    $N^{3}_v$ & Bounds &$N_x\times N_y\times N_z\times N_v^{3}$ &  $N_{\text{cycle}}$  &  $\text{T}$ (s)  & $T_{\text{cycle}}$ (s) & $T_{\text{cell}}$ (s) & (MB) \\
    \hline
    \hline
    \multirow{4}{*}{ $13^3$} &
    \multirow{4}{*}{\begin{sideways} $[-15,15]$  \end{sideways}}
    &$25^{3}\times 13^3$  & $32$   & $346$s   & $10.81$  & $6.92\times 10^{-4}$ & $2.4$\\
    &&$=3.4328125\times 10^{6}$ & & ($5.76$mn) & & &\\
    \cline{3-8}
    &&$50^{3} \times 13^3$  & $81$   & $4900$s   & $60.50$  & $4.84\times 10^{-4}$ & $15.5$\\
    &&$=274.625000\times 10^{6}$ & & ($1.36$h) & & &\\
    \cline{3-8}
    &&$100 \times 13^{3}$ & $160$  & $85720$s   & $535.75$ & $5.36\times 10^{-4}$ &  $115.5$ \\
    &&$=2.1970\times 10^9$ & & ($23.8$h) &  & &\\
    \cline{3-8}
    \multicolumn{2}{|c|}{ \textit{extrapol.} } &
      $200 \times 13^{3}$ & $320$ & $\sim 1.4 \times 10^6$s&  $\sim4400$ & $5.5\times 10^{-4}$& $\sim900$\\
    &&$= 1.7576\times 10^{10}$ & & ($16$d) & & &\\
    \hline
    \hline
    \multirow{4}{*}{$12^3$} &
    \multirow{4}{*}{\begin{sideways} $[-10,10]$  \end{sideways}}
    &$25^{3} \times 12^{3}$  & $27$   & $218$s   & $8.07$  & $5.17\times 10^{-4}$ & $2.3$\\
    &&$=27\times 10^{6}$ & & ($3.63$mn) & & &\\
    \cline{3-8}
    &&$50^{3} \times 12^{3}$  & $54$   & $2702$s   & $50.03$  & $4.00\times 10^{-4}$ & $15.4$\\
    &&$=125\times 10^3$ & & ($45$mn) & & &\\
    \cline{3-8}
    &&$100^{3} \times 12^{3}$ & $107$  & $38069$s   & $355.79$ & $ 3.56 \times 10^{-4}$ & $115.4$ \\
    &&$=1.728\times 10^{9}$ & & ($10.57$h) &  &  &\\
    \cline{3-8}
    \multicolumn{2}{|c|}{ \textit{extrapol.} } &
      $200^{3} \times 12^{3}$ & $214$ & $\sim 633440$s&  $\sim 2960$ & $3.7 \times 10^{-4}$& $\sim 900$\\
    &&$=1.3284 \times 10^{10}$ & & ($7$d) & & &\\
    \hline
  \end{tabular}
  \caption{ \label{tab:CPU_Sod3D_bis}
    3D Sod shock tube.
    The time per cycle is obtained by $T_{\text{cycle}} = \text{T}/N_{\text{cycle}}$ and
    the time per cycle per cell by $T_{\text{cell}} = \text{T}/N_{\text{cycle}}/N_c$.
    The lines marked with \textit{extrapol.} have been extrapolated by fixing $N_c$, $N_{\text{cycle}}$
    and $T_{\text{cell}}$.
  }
\end{table}
In Figure~\ref{fig:sod3D_CPU} we plot the CPU time (red or blue
symbols for each configuration and mesh points of the velocity
space) and the extrapolation curves $CPU(N_x,N_c,T_{\text{cell}}) =
\frac{N_{\text{cycle}}}{N_x} N_c T_{\text{cell}}$ for the 3D Sod
problem up to time $t_{\text{final}}=0.1$ for single processor
laptop computation on a fixed mesh in velocity space of $N_v=12^3$
points. We deduced that the FKS method can be used at most on a
single processor machine up to a $200\times 200\times 200$ cells for
roughly one week of computation. One also notices that the CPU time
linearly scales on a log/log graph as expected (right panel of
Figure~\ref{fig:sod3D_CPU})

\section{Conclusions}
\label{sec_conclu} In this work we have presented a new super
efficient numerical method for solving kinetic equations. The method
is based on a splitting between the collision and the transport
terms. The collision part is solved on a grid while the transport
linear part is solved exactly by following the characteristics
backward in time. The key point is that, conversely to
semi-Lagrangian methods, we do not need to reconstruct the
distribution function at each time step. In this first paper, we
have presented the basic formulation of this new method for the BGK
equation: Uniform meshes, piecewise constant discretization of the
velocity space and a simple projection towards the equilibrium
distribution have been considered.

The numerical results show that the method is incredibly fast. We
are now able to perform numerical simulations of the full six
dimensional kinetic equation on a single processor machine in several
hours. This important result opens the gate to extensive realistic
numerical simulations of far from equilibrium physical models.
Concerning the precision of the method, we observed, as expected,
that the fast kinetic scheme (FKS) is more dissipative close to the
fluid regime and very precise for gases far from the thermodynamical
equilibrium.

In the future we would like to extend the method to non uniform
meshes, more advanced boundary conditions and different
discretization of the velocity space. One expects with this last
point to increase the accuracy of the schemes without losing its
attractive efficiency. To avoid the loss of accuracy close to the
fluid limit, we want to couple the FKS method to an high order
solver for the system of equations which describes the fluid limit.
Finally, we want to extend the method to other kinetic equations as
the Boltzmann or the Vlasov equation.

%
%


\end{document}